\documentclass[aps,preprint,amsmath,amssymb]{revtex4}
\usepackage{amsmath,amssymb}
\usepackage{graphicx}
\usepackage{booktabs}
\usepackage[usenames, dvipsnames]{color}
\usepackage{multirow}
\usepackage{setspace}
\usepackage{rotating}
\usepackage{afterpage}
\usepackage[ pdftex, plainpages = false, pdfpagelabels, 
                 pdfpagelayout = useoutlines,
                 bookmarks,
                 bookmarksopen = true,
                 bookmarksnumbered = true,
                 breaklinks = true,
                 linktocpage=all,
                 pagebackref=false,
                 colorlinks = true,
                 linkcolor = BrickRed,
                 urlcolor  = blue,
                 citecolor = BrickRed,
                 anchorcolor = green,
                 hyperindex = true,
                 hyperfigures
                 ]{hyperref} 

\newenvironment{sectionsummary} {\begin{quote}\singlespacing } {\end{quote}}

\begin{document}

\title{A Complex Systems Science Approach to \linebreak Healthcare Costs and Quality}
\author{\href{http://necsi.edu/faculty/bar-yam.html} {Yaneer Bar-\!Yam}}
\affiliation{with}
\author{ Shlomiya Bar-\!Yam, Karla Z. Bertrand, Nancy Cohen, Alexander S. Gard-Murray, Helen P. Harte, and Luci Leykum
}
\affiliation{\href{http://www.necsi.edu} {New England Complex Systems Institute} \\ 
238 Main St.~Suite 319 Cambridge MA 02142, USA \vspace{2ex}}

\date{September 7, 2010; released June 28, 2012}

\begin{abstract}
There is a mounting crisis in delivering affordable healthcare in the US. For decades, key decision makers in the public and private sectors have considered cost-effectiveness in healthcare a top priority. Their actions have focused on putting a limit on fees, services, or care options. However, they have met with limited success as costs have increased rapidly while the quality isn't commensurate with the high costs. A new approach is needed. Here we provide eight scientifically-based steps for improving the healthcare system. The core of the approach is promoting the best use of resources by matching the people and organization to the tasks they are good at, and providing the right incentive structure. Harnessing costs need not mean sacrificing quality. Quality service and low costs can be achieved by making sure the right people and the right organizations deliver services. As an example, the frequent use of emergency rooms for non-emergency care demonstrates the waste of resources of highly capable individuals and facilities resulting in high costs and ineffective care. Neither free markets nor managed care guarantees the best use of resources. A different oversight system is needed to promote the right incentives. Unlike managed care, effective oversight must not interfere with the performance of care. Otherwise, cost control only makes care more cumbersome. The eight steps we propose are designed to dramatically improve the effectiveness of the healthcare system, both for those who receive services and those who provide them.
\end{abstract}

\maketitle

\section*{INTRODUCTION}
\label{sec:intro}

The US healthcare system suffers from high costs and low quality compared to healthcare systems internationally [1,2,3], as measured by reported life expectancy [4] and infant mortality [5]. High rates of nosocomial infection (infections acquired in healthcare settings) as well as adverse drug effects (errors in the administration of medication) manifest the need for improvement in the system of care. At a cost of \$2.5 trillion annually [6] the system is not delivering affordable, effective care. 
The paradox of higher costs and lower quality makes clear the existence of a systemic problem. How can we fix it? Complex systems science provides tools to address this question directly. In this paper we provide eight scientifically-based steps toward reducing costs and improving quality. Our suggestions arise from an analysis of the US healthcare system in particular, but they are broadly applicable when adapted appropriately.           

The eight steps are:
\begin{enumerate}
\item Separate simple care from complex care.
\item Empower workgroup competition as an incentive, and avoid regulating costs or quality. 
\item Create superdoctor teams to rapidly diagnose and treat highly complex conditions. 
\item Accelerate intake routing to rapidly identify the right provider.
\item Add redundancy to improve communication to prevent prescription errors.
\item Create disinfection gateways at spatial boundaries to reduce hospital-based infections.
\item Use e-records for research to supplement clinical studies.
\item Promote ``First Day''	 celebrations to encourage healthy behavior.
\end{enumerate}

Additional reading is available in the references provided at the end of the paper.

\section{Separate Simple Care}
\label{sec:one}

\begin{sectionsummary}
 {\bf Scientific principle---Matching complexity and scale:} For an organization to perform tasks effectively, it must be organized so that it can match both the scale (rate of repetition) and complexity (variety) of those tasks. When large scale tasks are performed by an organization designed for complex tasks, the result is inefficiency. When high complexity tasks are assigned to an organization designed for large scale tasks the result is non-optimal (wrong) acts, i.e. errors. Separating tasks by scale and complexity enables simple, mass-applicable care to be performed by individuals and organizations (retail clinics in this case) well suited to those tasks, and complex tasks to be performed by individuals and organizations  (physician practices) designed for those tasks.
\end{sectionsummary}

Healthcare work may be divided into two types: simple care, which is the same for many people, and complex care, which is different for each individual. 
	
Simple care includes preventive services, such as health screenings, vaccinations, and healthy-habits counseling sessions. Complex care includes the individualized diagnosis and design of treatment. 
	
Physicians are specially trained to diagnose and treat complex medical conditions. Nonetheless, one finds that physicians and their offices are typically responsible for simple, standard care in addition to complex, individualized care.

That poses a problem. Asking the same organizational structure to provide mass-applicable preventive care and complex individual care is like asking an expert violin craftsman to provide all the chairs for a new concert hall. The mismatch between the organization and the task leads to ineffectiveness and inefficiency.

Ironically, instead of streamlining the delivery of high-volume simple services, most cost-reduction efforts to date have tried to make complex tasks simpler and faster. Industrial-style efficiency is poorly applicable to doctors' diagnoses and treatment of individual patients, however. Trying to speed and simplify doctors' work assembly-line style reduces doctors' time to make complex decisions, which is not a good idea if we want doctors to be careful and make the best decisions possible. At the same time, many healthy patients are receiving insufficient preventive care, since doctors are being asked to provide many of these services. The volume of preventive care needed is too great for the current system to handle it effectively.

What can be done? The solution is to separate the tasks. Let doctors perform the complex tasks that they do well, and delegate preventive-care tasks such as vaccinations to an organization suited for simple, repetitive tasks.

In many hospitals and doctors' offices, simpler tasks such as drawing blood and taking x-rays are performed by professionals trained for these specific, frequent tasks. This idea can be applied much more broadly.

We can improve the healthcare system dramatically by separating the simple services that many healthy people need even further, delegating them not just to different individuals but to different organizations.

We are beginning to see this concept in programs that make flu shots available in supermarkets and airports, and in the growing number of Òretail clinics.Ó

Since retail outlets at malls and supermarkets serve many people with similar needs, Òretail clinicsÓ make sense; they can readily provide routine and preventive care such as health screenings, vaccinations, and dissemination of public health information. These clinics have the additional advantage of locating preventive services where healthy people frequently go, rather than requiring them to make less convenient trips to their physician's office.

CVS is installing MinuteClinics in its pharmacies. Walmart has such clinics at over 50 locations and is planning thousands. These clinics, originally developed to provide routine treatment for minor problems such as strep infections, now also offer preventive services, including vaccinations, cholesterol and other tests, and school physicals.

What is the payback in widespread retail clinics? To be sure, a retail setting offers convenience and efficiency in implementing preventive care via large volume, simplicity, and a focus on healthy people. Whenever a large number of similar tasks are to be performed, the medical system is well-served by moving such care from physicians' offices or hospitals to the retail setting.

Besides easing the burden on doctors---freeing them for complex tasks for which their time is now too limited---the separating-out of mass care from individual care would streamline high- volume processes. This would address the excess costs that arise when the performer and the task don't match.

More people would gain access to routine care due to the convenience of location and avoidance of the need to make appointments, travel to the doctor, and wait.

What is clear is that making preventive healthcare more accessible can reduce illness, further easing the burden on the medical system. It is time for the medical and insurance communities to embrace this solution.

\section{Empower Workgroup Competition}
\label{sec:two}

\begin{sectionsummary}
{\bf Scientific principle---Evolving complex systems:} Highly complex systems arise through evolutionary change that involves replication, variation and selection. We can understand this in social learning as mimicry (replication), creativity (variation) and competition based on comparative evaluation, i.e. performance feedback (selection). Highly complex tasks cannot be performed by a single individual (they exceed an individual's complexity). It is not even possible for an individual to manage or design a system to meet the challenge of high complexity. Instead evolutionary processes that provide for feedback and learning of the group performing the task enable progressive improvement. Just as team competition drives improvement in sports, competition between teams of care providers enables improvement of outcome measures. 
\end{sectionsummary}

The current use of ``managed care'' to improve medical performance and reduce costs is inherently flawed. To understand where the flaws lie, it's useful to ask: What roles should government, management and practitioners play in healthcare and in healthcare decision-making? Today, they often serve in the wrong roles. That's because they serve within the wrong system structures. Neither traditional centralized management nor free market competition scenarios work for healthcare and yet those settings are perceived as the only two options.

Whether by government or by insurer, efforts to control healthcare costs are typically deployed within a setting of centralized management to dictate how to allocate limited funds. Yet healthcare is a highly complex system, and, as we see in the failure of the USSR and other centralized economies, centralized control doesn't work for complex systems.

At the same time, free-market competition, which may result in rapid improvements for, say, the electronics industry, is ill-suited for managing healthcare. After all, patients generally can't shop around for the best hospital.

Are these the only two options?

There is a third: empowering workgroup competition. Groups of care providers who together can be responsible for medical outcomes and other performance metrics compete against other groups in the same hospital and between hospitals. Workgroups must become teams in a peer competition to improve care. This team competition approach combines the best aspects of both the free market and centralized management; it allows the spirit of competition to spur advances and improve performance, while still allowing management to set objectives.

In free-market competition, the goal is financial gain. In workgroup competition, the objective is to be a top performer according to carefully designed metrics that measure both cost and quality. This kind of competition works for sports teams, students competing for grades, and in other competitions where the goal isn't just to make money.

In order for medical care to improve, the people engaged in providing that care, who know the most about what to do, must be the ones who have control over care decisions, and must be the ones with responsibility for outcomes. However, performance should not be measured at the level of individual doctors and nurses, because outcomes often rely on an entire workgroup's performance---e.g., how nurses or physicians communicate information across a shift change has a huge impact on outcomes, and communication relies on how people work together.

How does a system empower the people who work together, and who can take on such responsibilities---and not tell them what to do? Through workgroup competition. Fostering workgroup competition means first identifying and solidifying groups that can be responsible for outcomes.
	Nurses at a nursing station responsible for the care of patients in a specific part of the hospital could be a workgroup, taking responsibility for improvement in areas like infection rates and patient satisfaction.
	In many hospitals, nurses, technicians, and an anesthesiologist are assigned to a surgery based on who happens to be available. In some, however, each surgeon has his or her own team of nurses and techs, and an anesthesiologist with whom he or she works. This team-based approach allows workgroup members to get to know each others' styles and to work smoothly together. A team approach also allows them to improve their outcomes as a group.
	One emergency room, in the Washington Hospital Center of Washington, DC, takes just such a team approach, dividing the emergency room staff into teams who care for a patient from entry to release or admission to the hospital.
	Workgroups must be of the right size and function to be able to improve based on the measures evaluated. Some efforts have attempted to produce competition between entire hospital systems, but these units are too large and unwieldy. They focus responsibility on the hospital system management. Since these efforts do not directly measure workgroup performance, they do not enable practitioners to take responsibility for their outcomes or work together to improve their performance. Having provider workgroups rather than management assume the responsibility for healthcare performance in a competitive environment is key.
	Once workgroups are created, the metrics of competition must be designed. Healthcare administrators should determine workgroup-performance measures that cover health outcomes, costs, lengths of hospital	stays and patient satisfaction. Remember---what is measured is what will be improved: performance metrics must be designed carefully, and should be revisited periodically for updates and improvements.

Finally, workgroups' performance on each of the measures should be publicized to all the groups on a regular basis---say, monthly. The workgroups would compare their results with those of other workgroups providing similar types of care, like other surgical teams or other nursing stations, and then be responsible for their own improvement.

Care must also be taken to set an appropriate tone for the competition. Just as with sports, rules must be established which will encourage good sportsmanship. Also, as teams improve their performance, swapping team members among them causes better strategies to be shared and improve the performance of all teams.

When workgroups' scores are in focus, both the scores and the performance they measure improve radically. Members of a workgroup will work together to improve their performance. They will innovate, and they will emulate improvements that they see others adopting.

Some hospital systems publicize their performance measures so that consumers can compare hospitals or systems. This won't improve performance, however, because patients can't always change providers, and because hospital systems don't have much control over their practitioners' performance.

Arranging hospital operations such that groups consistently work together, and then identifying these workgroups and entering them into a spirited, friendly competition based on well-designed performance metrics will lead to dramatic improvements in results.

Besides improving performance on the measures evaluated, workgroup competition will change the way efforts to control healthcare costs are directed. We will move from a limited focus on cutting healthcare costs to a broader focus on improving the healthcare system at lower cost.

 {\bf Why centralized management doesn't work in healthcare:}
\begin{itemize}
\item Decisions about care should be made by the people who understand the need for care, the healthcare providers.
\item The decisions must be made on a case-by-case basis. Abstractions according to generalized rules degrade the effectiveness of care.
\end{itemize}
 {\bf Why free market competition doesn't work in healthcare:}
\begin{itemize}
\item The consumer has limited knowledge of how well each hospital or doctor's practice works. 
\item The ability to choose is often limited by emergency circumstances, capacity of the provider, insurance policies, or restrictions on changing providers.
\end{itemize}
 {\bf Why team competition works:}
\begin{itemize}
\item Professionals care about comparisons with their peers. 
\item They have the most knowledge and ability to improve outcomes. 
\item Team competition, like competition in biological evolution, is the natural way to improve performance.
\end{itemize}

\section{Create Superdoctor Teams}
\label{sec:three}

\begin{sectionsummary}
{\bf Scientific principle---Organizing for complexity:} Systems must be organized for the complexity of the tasks that are to be performed. Specialization enables a group of individuals to perform more complex tasks than an individual can do alone. It does so by routing different tasks to different individuals, enabling them to perform a wider variety of tasks. The number of distinct tasks that can be performed by the system of specialists grows linearly with the number of individuals (it is the sum of the number of types of tasks each individual can perform). However, a collaborative team enables each individual to contribute a different dimension to the task performed by the group, so that the number of types of tasks can be as high as the product of the number of tasks each individual can perform. As there are increasing number of specializations, and conditions that require multiple specialists to address, this implies that we have reached the point in complexity of care that teams are increasingly necessary to address complexity of medical care. 
\end{sectionsummary}

A hundred years ago, physicians were generalists, treating most medical conditions. Humanity didn't have nearly as much medical knowledge and knowhow back then so that for the most part a single doctor could master what was known. That has changed.

Medical knowledge now far exceeds a single expert's ability to master it. Medical students receive a general training and then they specialize, seeking to learn just one small piece of what we know about medicine.

\begin{figure}[h]
\includegraphics{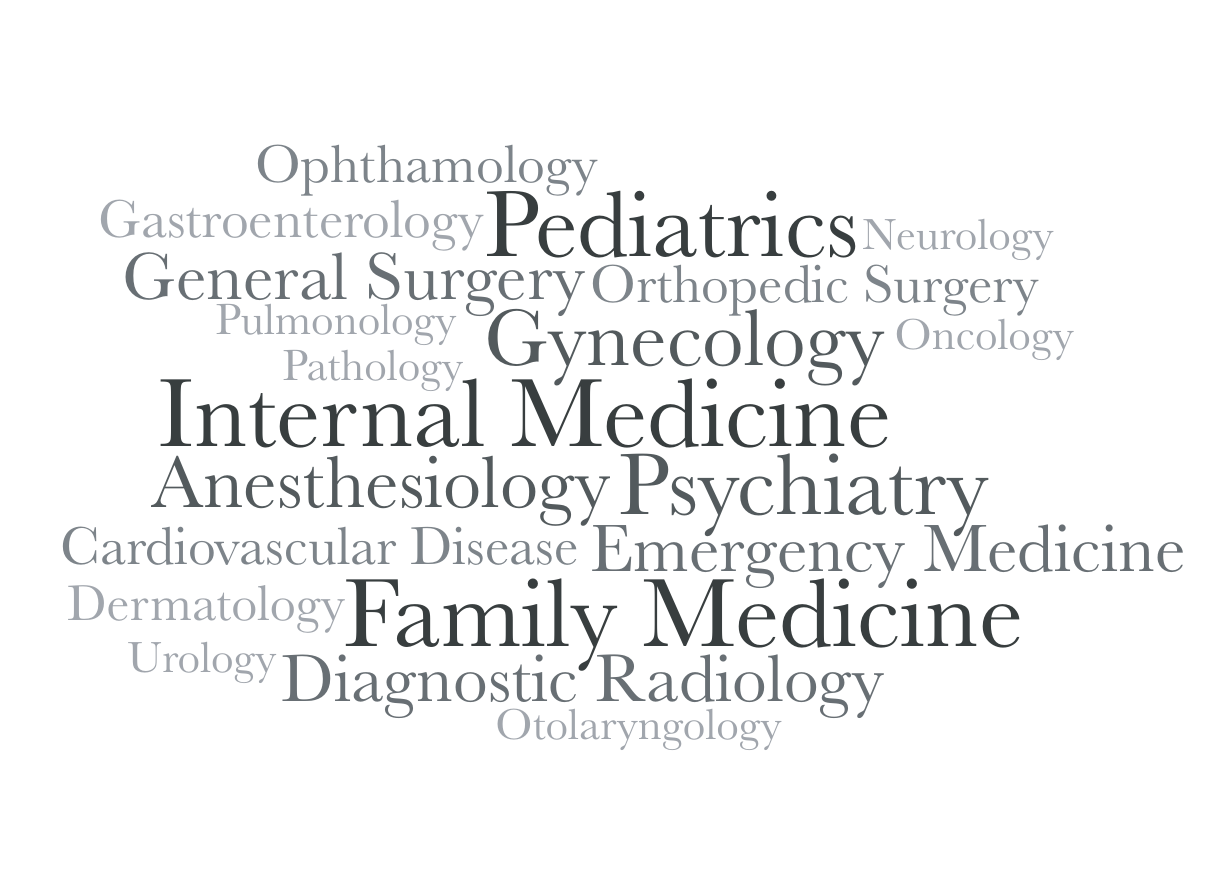}
\caption{\label{fig:specialization}Specialization reflects the growth in medical knowledge, which is much greater than an individual's capacity to master.}
\end{figure}

Specialists have become essential because of the complexity of care. The more we learn, the more kinds of specialists are needed. Increasingly, however, it is necessary to have patients see multiple specialists for a single problem, which causes fragmentation and delays the care. Furthermore---and critically---the interplay between multiple causes of a single condition, or multiple aspects of its treatment, makes it difficult for the separated specialists to address such complex problems.

What is the solution?

A human being is a single working system and specialists must be able to work together as an integrated unit for diagnosis and treatment. Specially constituted teams of physicians and other care providers who work together on a regular basis should address the more complex problems. The cost of having such a team in place might seem high, but for complex cases such a team will prove to be more effective and less costly than the alternative---the difficulties, delays, and costs inherent in multiple appointments. The challenge is making sure the teams can work together smoothly and efficiently, and with better results than specialists working separately.

A well-integrated team of specialist physicians can be thought of as a ``superdoctor.'' In order for medical teams to be superdoctors, they must get to know each other's strengths and styles and act together seamlessly. Well- integrated teams have the combined specialized knowledge of each member and more: they have the ability to relate these different domains of knowledge and combine them in new ways. Moreover, they can act rapidly with this combined knowledge. They can be an important part of the solution to the problems of fragmentation.

\begin{figure}[tb]
\includegraphics[width=0.9\textwidth]{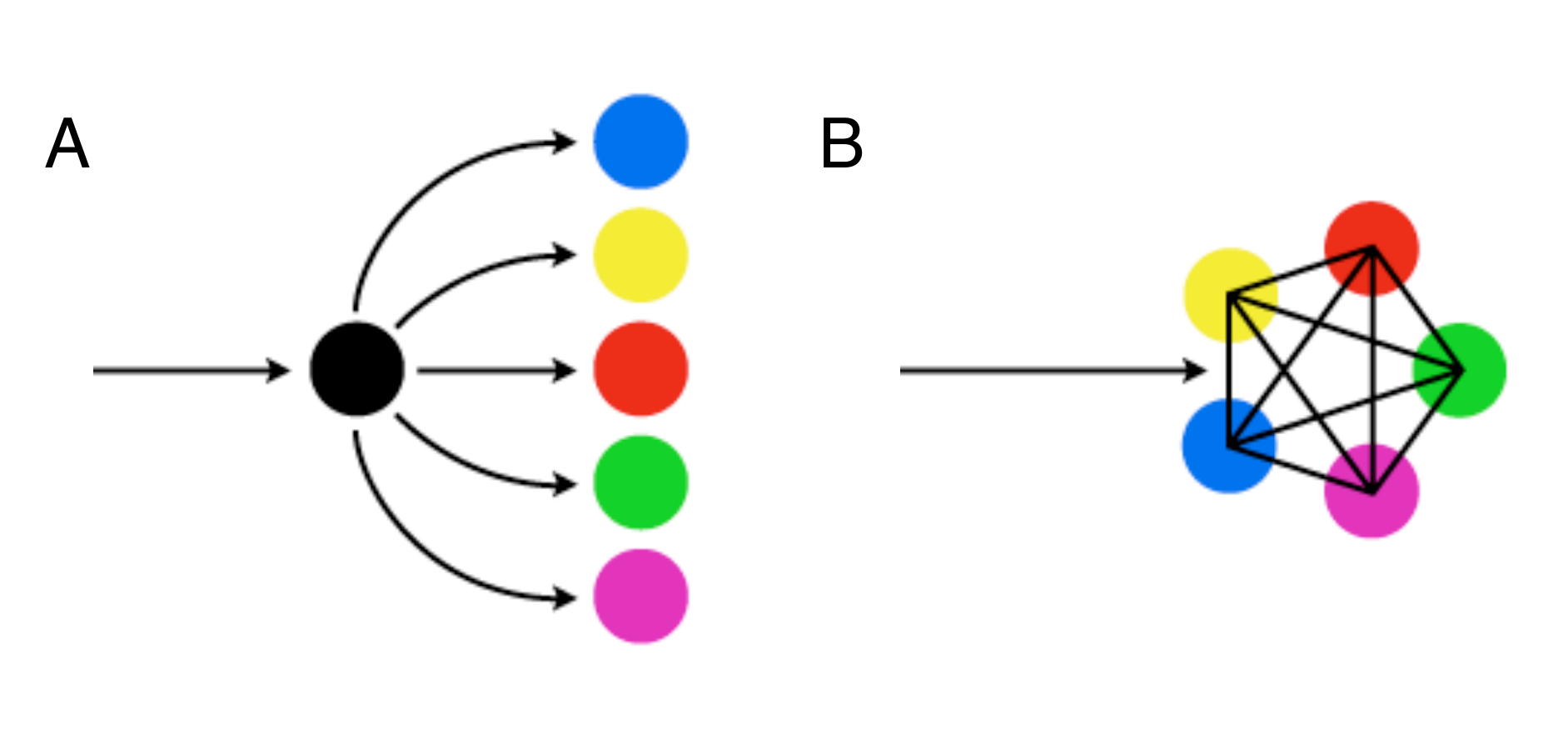}
\caption{\label{fig:complexrouting} A. The knowledge of specialists is brought to bear using a routing system. B. When conditions require multiple specialists, multidisciplinary teams can provide integrated comprehensive care.}
\end{figure}

\begin{figure}[tb]
\includegraphics[width=0.9\textwidth]{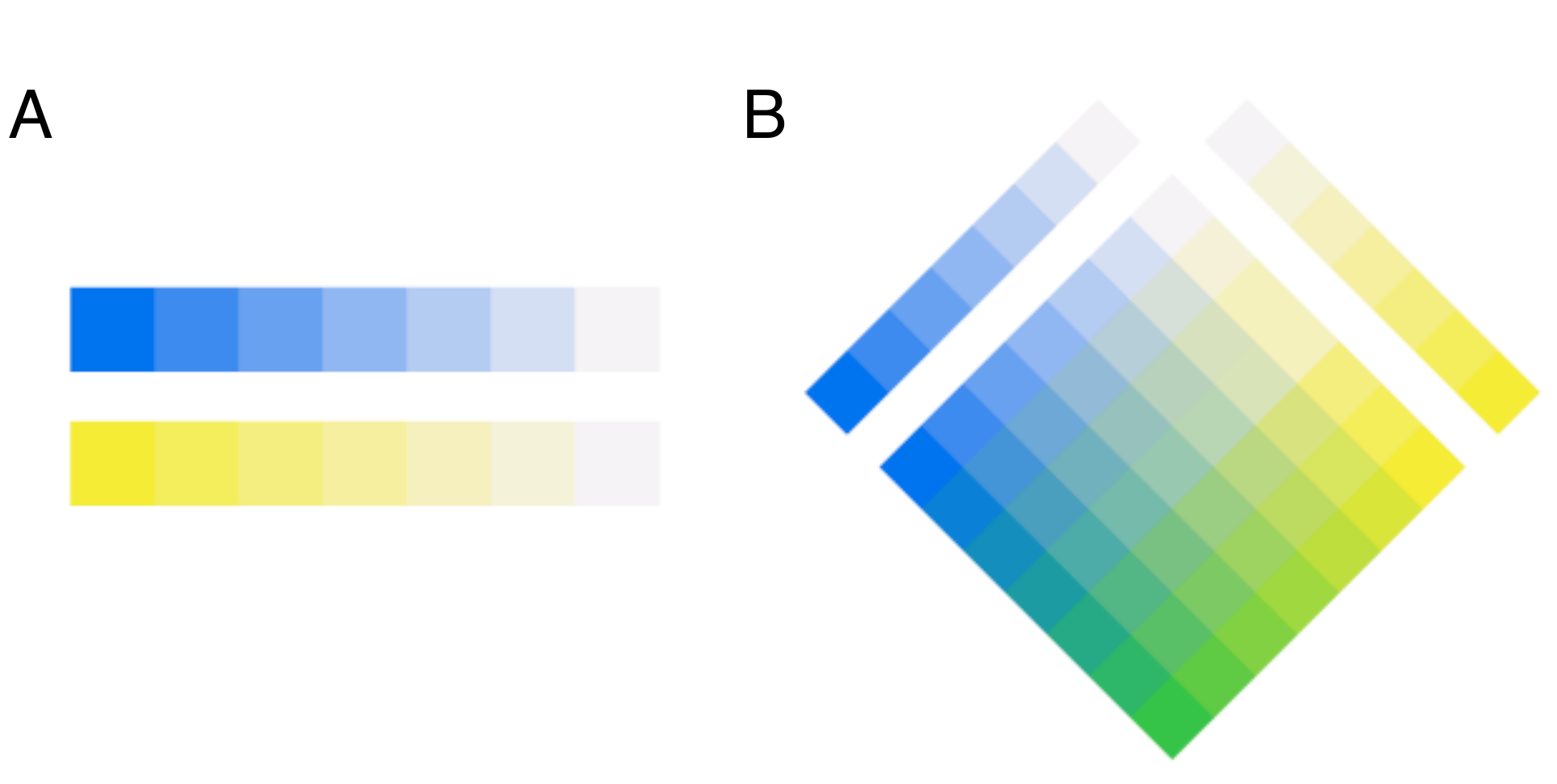}
\caption{\label{fig:teamadvantage} Distinct tones symbolize the distinct tasks possible for an individual or team to perform.  A. By separating distinct types of tasks, specialists can address many more conditions than a single practitioner. B. Through joint action, teams can address an even more diverse set of conditions than a similar number of specialists. The case shown is for two specialists and two-member teams.}
\end{figure}

Such teams have become standard practice in cancer care, where specialists in imaging, surgery, radiation therapy, and chemotherapy often meet and work together to treat patients. The wide diversity of cancers and of individual responses to treatment make the team approach necessary for effective care. These teams generally also include non-physician practitioners. While the team approach is most widely used for cancer, some medical centers, recognizing the problem of fragmentation in care, are using the team approach for other conditions.

To be most effective, superdoctor teams need to work together on a regular basis. If you were to throw together several sports players---even professional athletes---to play as a team without training together, they would not play as well as they would with team members who they were used to. Similarly, medical teams must ``practice'' together to fully leverage their collective ability.

We can take clues for the formation of superdoctor teams from the types of cases that currently require many specialist appointments---teams should be formed that can handle these cases together more effectively than the specialists could working separately. The advantage of the team is not just the ability to do what the individual specialists would do separately; it's the ability to treat a wide range of conditions effectively, to make very subtle distinctions that are important for effective care, to solve the cases that are the most difficult due to the interplay of multiple causes or complications.

The capabilities of superdoctor teams will grow through being challenged, and they will learn from experience. Innovation in their composition and testing their abilities is key. We can only discover their effectiveness through observing how they respond to challenges. Measuring their effectiveness brings us back to Step II.

Superdoctor teams can assume the dynamics we described in Step II: Empower Workgroup Competition, competing against one another and continuously pushing the boundaries to improve care and reduce costs. By measuring their performance, we can learn how to build more and more effective teams, both in terms of choosing types of specialists to be on a team and in their specific interactions.

It is important to note that, in trying to stem the cost of specialist care, alternative cost-cutting approaches have been tried but have not been successful.

Some have proposed having primary care physicians treat more cases, to reduce the number of specialists that patients see as a way of reducing healthcare costs. This approach, though at times politically popular, is ineffective. Family physicians can treat a certain set of conditions, but they do not have the specific knowledge to treat many complex, more specialized conditions.

Of course, we will still need primary care physicians---many problems are best treated by a single person knowledgeable about a wide range of conditions. We also will continue to benefit from individual specialists, or from specialists who don't normally work together collaborating for particular patients. This works fine for problems of intermediate complexity.

However, for the increasing number of highly complex cases, superdoctor teams are necessary for comprehensive, integrated, cost-effective, quality care.

We must take steps to form innovative specialist teams that can treat the most complex cases successfully and cost-effectively. Introducing such teams is essential if we are to put our vast medical knowledge to effective use.

\section{Accelerate Intake Routing}
\label{sec:four}

\begin{sectionsummary}
{\bf Scientific principle---Limiting rates for dynamic response:} The rate of response of a system is a key aspect of its ability to perform time sensitive tasks (such as medical care). A system only responds as well as its rate limiting step, i.e. the step that takes the longest time to perform. The current design of the US healthcare system involves the use of ``gatekeepers'' who serve to route individuals to the care they need. Evidence suggests that this task has become the rate limiting step. The gatekeepers become overloaded, wait times to see them are extended and the routing process itself often involves many time-delayed steps. Improving the system behavior involves: explicitly recognizing response time as a critical dimension of care, identifying the time of first contact rather than the time of intake as the beginning of the medical response process, expanding the set of gatekeepers to allow adequate parallel intake channels, and improving the routing function to better utilize available channels of care that work in parallel. These interventions also should reduce the complexity of the initial intake process.
\end{sectionsummary}

One of the strengths of our current healthcare system is that the critical task of routing patients to appropriate specialists is performed by primary care physicians. They are also often on call to respond to inquiries about urgent care or referral to an emergency room.

How patients are routed to appropriate care is a crucial aspect of running an efficient medical system. As we consider changes to our healthcare system, it is vital to note the importance of accurate and timely routing, to focus resources on keeping this process reliable, and to improve upon it where possible. Routing is known in medical circles as triage. Unlike the triage in disaster or wartime, triage in a modern medical setting simply refers to the act of directing patients to the appropriate care.

Today, most routing occurs in the primary care physician's office or through the physician's after hours call-in system. This process often works smoothly, with knowledgeable family physicians, internists or pediatricians assessing patients in a timely manner and routing them appropriately. They may treat a patient directly or refer to an appropriate specialist or clinic.

When care is required after-hours and the PCP is not available, patients call a service to reach the doctor on call. Some medical offices provide extended-hour urgent visits. Some practices and insurance agencies also provide 24-hour phone access to medical professionals through a ``nurse line.'' These call-in services can provide feedback to the patient as to whether he or she should go to the emergency room, or whether their condition can wait.

However, the medical routing system doesn't always work efficiently.

Individuals who are uninsured, or who do not have a primary care provider, often resort to the emergency room for non-emergency care. Without a PCP to act as a router, and without access to a 24-hour phone support, these individuals' options are limited. If their medical problem occurs at night, or if they can't make it to a walk-in clinic during the clinic's hours because of work, childcare, or other responsibilities, the emergency room becomes the only option. This puts a huge strain on the emergency care system, since a large portion of its resources must be devoted to treating or routing these cases rather than on the truly urgent ones.

Even for people with insurance and a PCP, the intake system often doesn't work well. The wait for an initial appointment with a family physician averages 20 days in the United States as a whole. According to a 2009 survey, the average wait time to see a family practitioner in Washington, D.C., was 30 days; in Los Angeles, it was 59 days and in Boston, 63 days.

This delay in the initial care and routing process indicates that something is amiss with the intake system. The influx of patients seeking initial evaluation and referral to specialists overtaxes the primary care system. What's more, delays in diagnosis can have serious health consequences.

The initial triage decision often falls to the receptionist, who is generally unprepared to properly make such decisions. When the doctor is too busy to see everyone who wishes to be seen, the receptionist is put in the position of deciding which patients must be seen urgently and which can wait for an appointment---an appointment that may be several weeks away.

No matter how good the care is once a patient gets to the right place, delays mean that the healthcare system isn't working well, not for quality of care, where delays may compromise the patient's health, and not for costs.

Even where the current routing system works well, it can be improved using advances in technology. The advent of email and other forms of electronic information and communication have changed our expectations of response time. The medical routing system should make use of these advances to build on the existing structure, speed response time and improve outcomes.

Perhaps the best way to think about an effective rapid-response system is ``Triage on Steroids''... or ``Super - 911.''

A routing service should be available 24 hours a day, seven days a week, and be performed by a knowledgeable and capable medical professional. This is the gold standard of medical routing. The system should be accessible via phone, Internet, and in person. Most of the traffic should be handled by phone or electronically; for in-person routing, the staffed routing site might be near to---but separate from---an emergency room, or perhaps at a pharmacy.

The first task of a routing service is to serve as a ``super-911,'' to identify emergencies and reassure if urgent care is not needed.

Second, if it is not an emergency, the intake specialist could determine the complexity of the response needed. Patients requiring a simple, standard response, especially preventive care such as flu shots, could be directed to a preventive care clinic (see Step I). Patients with more complex problems should be directed to their primary care provider or, where the determination can be readily made and the primary care provider is otherwise overburdened, an appropriate specialist. Finally, patients with especially complex problems might be referred to a superdoctor team (see Step III).

\begin{figure}[h]
\includegraphics[width=.6\textwidth]{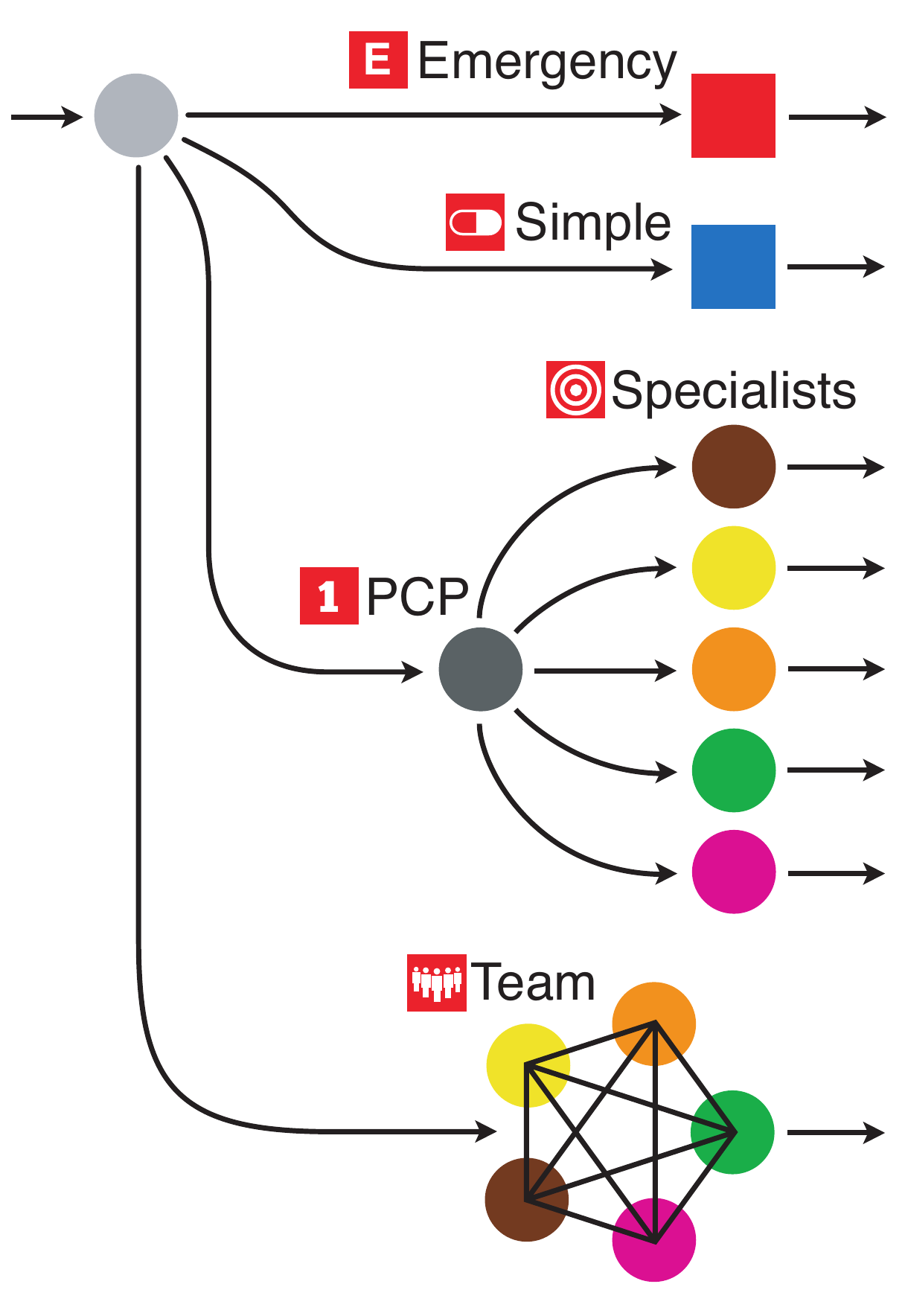}
\caption{\label{pathway} A schematic of an intake routing system.}
\end{figure}

To accelerate the appointments for an initial evaluation, some family practices are adopting a system known as ``open access,'' designed to facilitate routing and expedite care. In an open access setting, no appointments (or only a limited number) are made significantly in advance. Instead, patients call into the office when they require care and are given an appointment that same day. This system provides an opportunity for a very rapid initial evaluation, allowing for routing decisions to be made literally hours after symptoms manifest.

There is another difficulty that could be addressed with creative use of new technology. Often, the best person to determine whether a particular specialist should be seen is the specialist himself or herself. But a patient moving from specialist to specialist to find out who should provide treatment is not a good strategy. It is inefficient and potentially costly in health consequences.
One approach to solving this problem is to use information routing rather than patient routing. The key is information-gathering and communication. Most of what happens at an initial medical visit to a clinic or primary care physician is a gathering of key information that will serve to determine which specialist should be seen. In information routing, after the initial visit, the information, not the patient, would be forwarded to a number of specialists.

The specialists could rapidly evaluate whether, based upon this limited information, they should be seeing the patient. Or, a specialist might provide a question---if the patient has such and so a symptom or such and so a test result, then they should be seeing the patient, e.g., ``If the patient's ears hurt while the other symptoms occur, she should see me. If not, I'm not the right specialist for this case.''

This information-based routing system, on the specialist level, serves patients better, and costs less, than patients being sent around to several specialist appointments in order to route them correctly.

An ``everywhere and always-on'' routing system could be made available instead of the more usual answering services, by primary care providers, provider systems or insurers. Such a system would relieve emergency rooms of having to perform the routing of non- urgent cases, freeing them to focus on the urgent and emergency care they are supposed to be providing. This type of routing system would also relieve some of the burden of PCPs, and shorten patients' wait times for routing and treatment significantly.

It is clear that an accessible and reliable 24/7 accelerated triage mechanism---staffed by intake specialists and augmented by an information-transfer system---will dramatically improve our medical system's cost-efficiency and ability to serve patients well. Wait time for care will be dramatically reduced, emergency rooms will be put to their proper use, and the burden on primary care providers will be lightened. Augmenting and improving our existing routing system is crucial to improving healthcare quality for all.

\section{Improve Communication}
\label{sec:five}

\begin{sectionsummary}
{\bf Scientific principle---Information theory, errors, and redundancy:} According to information theory, adding redundancy to a message dramatically reduces the error rate when it is transmitted through noisy channels. Human-to-human communication can be analyzed as occurring through a communication channel, whether oral, handwritten or through electronic means. Information theory uniquely identifies the origin of errors and how to alleviate them. Errors in the medical system that result from miscommunication can be analyzed using this approach and the introduction of the necessary redundancy can be used to drastically reduce medical errors. Prescription miscommunication has been documented as a major cause of loss of life and other adverse outcomes and should be addressed in this way. It is a common misconception that automation (electronic prescription systems) reduces error rates. This is incorrect, unless such systems introduce the necessary redundancy specified by information theory.
\end{sectionsummary}

Ten years ago, the Institute of Medicine's report on the extent of serious medical errors brought the issue to the attention of medical professionals and the public. According to the Institute and the FDA, medication-related errors cause over 1 million harmful drug events each year. Even one case of medical error may result in tragedy for those directly affected and may be traumatic for the professionals involved.

How can this problem be solved?

First, it must be said that the often-suggested electronic prescription system is not the solution to medication errors---unless the system is well designed. Research shows that different electronic systems affect errors quite differently, ranging from eliminating 99\% of them to increasing the error rate and all possibilities in between. Moreover, these systems can cause a variety of unanticipated side effects that compromise patient safety. This paradox can be understood once the real sources of medical errors are understood.

For many errors, the solution lies in adding redundancy. What does this mean?

To explain, we can turn to another context where the prevention of errors is important: writing checks.

Where money is involved, we are careful to make sure the information is conveyed clearly. To this end, we write the amount twice, in both words and numerals. This is done, purely and simply, to prevent errors. Electronic check-writing systems also make sure that critical information is ``double-checked.'' Another example is the double entry of e-mail addresses or passwords when one registers for online accounts. Why enter the same information twice? To make sure it is correctly received.

The same principle of redundancy can and should be applied to writing prescriptions.

Why isn't this done already? The system we use today for writing prescriptions was developed when there were far fewer medications. As the number of possibilities increases we have to be increasingly careful to make sure that enough information is communicated so that the right prescription is delivered.

Thus, whether written or electronic, what matters is how well the system is designed.

The caveat is that every critical piece of a prescription must be written twice, to ensure that few if any errors occur. There are five critical pieces of information on a standard prescription: Patient, Drug, Dose, Route (oral, intravenous, etc.) and Time. Each of these must be written in two ways, or double-checked after electronic entry.

For example, the patient could be designated by both name and ID number. The drug could be doubly specified by writing both the medication and the indication (the condition for which it is prescribed), or both the generic and trade names. Dosage, route of drug delivery, and time of administration could be written out fully and abbreviated, rather than given only in abbreviated form.

For electronic systems, auto-completion and simple check boxes should be avoided. These items are more prone to error precisely because they are quick and easy. Instead, it is important to have the prescriber provide all key information longhand and verify it. Writing something twice admittedly takes more time but the prevention of errors, as in writing checks, must be considered of primary importance.

\begin{figure}[h]
\centerline{
\includegraphics{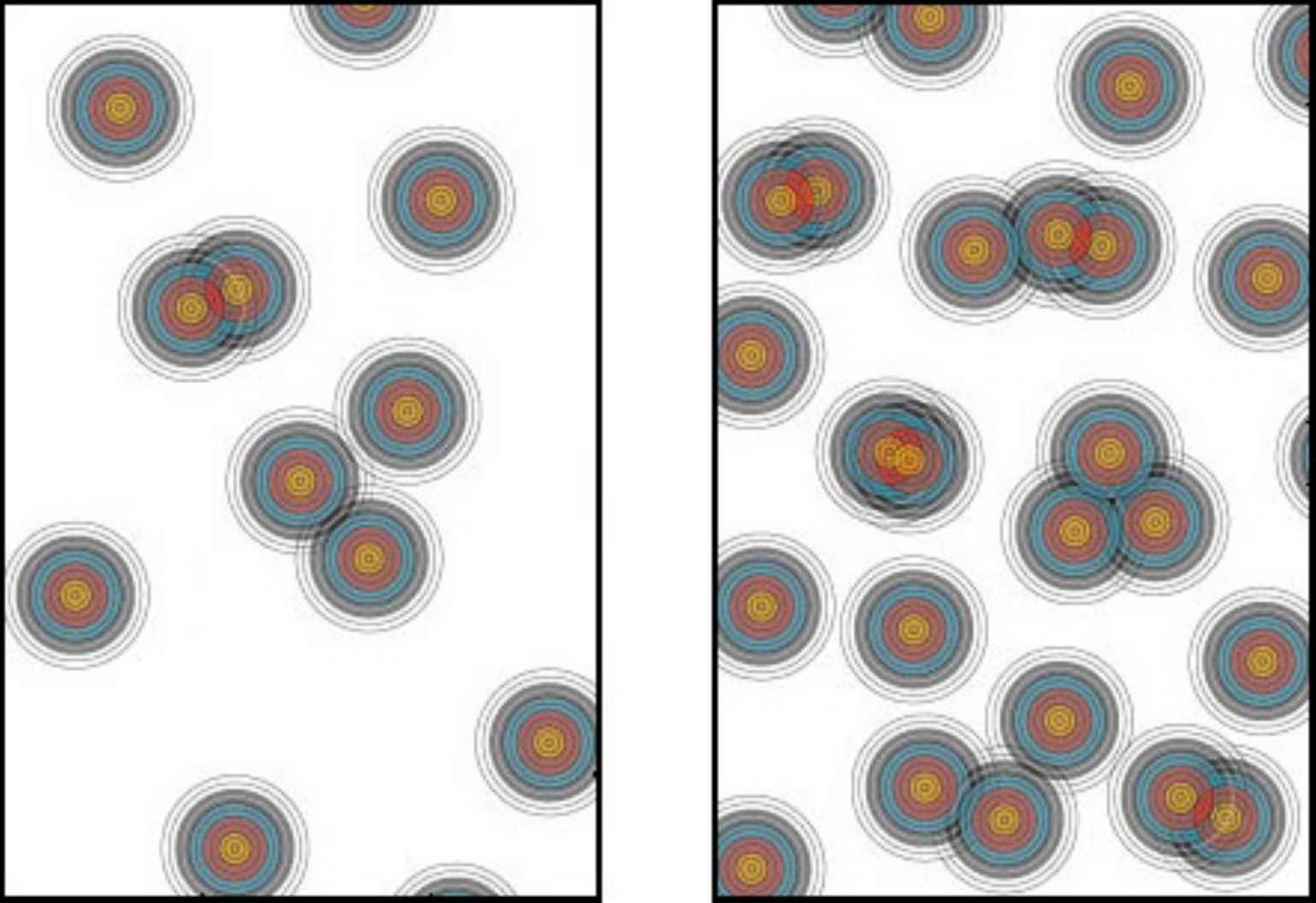}
}
\caption{\label{fig:targets}Hitting the right target is harder when there are more targets. The more medications and treatments there are, the more accurate the system has to be to avoid errors that shift from one of the possible medications to another.}
\end{figure}

It is possible to write less when there is less potential for misunderstanding. For example, if the route is already determined by the medication, then the route can just be indicated as ``Standard.'' For now, however, we should be conservative in shortcuts; once medication errors are dramatically reduced, we can carefully study which efficiencies can be implemented without errors being introduced.

Electronic systems also should be carefully designed to avoid distraction and disruption. The difference between a well-designed intuitive way of entering prescriptions, with appropriate redundancy, and a poorly designed system is the difference between success and failure.

Communication is not only central to prescription errors, but is also central to other forms of medical errors. Errors generally arise not because of an individual's action, but because of the way individuals work together. Improvement of communication and coordination is often the solution. The development and competition of workgroup teams recommended in Step II are key to reducing errors throughout healthcare, because such groups can improve local team coordination and communication.

In looking for ways to solve the problem of medical errors, improving upon the analysis of the source of medical errors is important too. Too often, the process of examination only looks at a specific error---what went wrong in this particular case---and a particular practice is blamed, and a particular solution is offered to that practice. Instead, we should abstract from the level of individual errors and find the patterns among effective and ineffective cases.

Focusing on just the individual error is as ineffective as a tennis player only practicing the one last shot he or she missed, over and over. In most cases, it makes more sense to work on improving speed, agility, and the player's ability to respond to a large set of possible shots. The next challenging shot will not be the same as the last one.

The same holds true for medical errors: understanding the many possible ways errors can occur rather than just the last one, and the way things work correctly, will allow us to recognize the weaknesses and improve the strengths of the system.

The Institute of Medicine originally reported up to 100,000 deaths per year due to medical errors, and up to \$29 billion in additional costs incurred. More recent reports have found these numbers to be even higher. The price in human and financial terms is too great. We can and must fix the problem.

\section{Create Disinfection Gateways}
\label{sec:six}

\begin{sectionsummary}
{\bf Scientific principle---Dynamics of well-connected versus modular networks:} In many systems the difference between a geographically partitioned and a well-mixed one is significant. In a well-connected system, the large number of pairwise interactions makes highly likely the spreading of any transmittable condition. Indeed, analysis shows that infections of high virulence and transmissibility only survive in highly connected systems. Understanding the flow of contagion through a system involves mapping out the set of contact points and the network of transmissions that result. Today the focus on reducing infection transmission in hospitals is on reducing the likelihood of transmission through each of the many individual contacts. However, because the system is highly connected, the probability of transmission is high even when there is a low individual contact transmission probability. Spatial, and more generally, hierarchical partitions are a powerful approach that reduces the overall connectivity of the system and thus dramatically decreases the sustainability of infections, inherently making the system non-conducive to highly virulent and transmissible strains.
\end{sectionsummary}

Infections acquired in hospitals, known as HAIs or nosocomial infections, are often resistant to antibiotics and thus particularly dangerous. Each year, the estimated 1.7 million infections cause nearly 100,000 deaths in the United States. Many patients in hospitals, nursing homes and clinics become sicker from these infections than they were before they sought care. These infections also play a significant role in costs---HAI hospital costs alone were recently estimated at between \$30 and \$45 billion.

Current recommendations for reducing hospital-acquired infections target the patient's immediate environment and interactions with care providers. Hand washing by care providers before and after patient contact is a key part of protocols in patient-focused transmission prevention. The wide variety of other measures include identifying patients who enter the hospital with infections for additional isolation, extra care to avoid catheter-associated infections, and augmented surface sanitation---of bed rails and controls, light switches, partition screens, faucet handles, and the like.

Collectively, recommended protocols have been shown to reduce transmission, are cost effective and could be more widely adopted. Still, the attention and effort involved are significant and progress in eliminating infections is slow.

Underlying the widespread prevalence and difficulty in addressing these infections is the large number of contacts between care providers and patients. Because there are so many contacts, the effort involved in making every contact safe is huge and this effort burdens already busy care providers.

How can we speed up progress?

We need to expand our view beyond the point of contact between patients and providers to think in terms of the overall process of transmission within a hospital and between care facilities.

Each transit across a boundary between domains should be considered as a potential ``transmission'' of pathogens that will infect a unit, ward, floor, building or care facility. At these boundaries, protocols of disinfection should be designed to reduce pathogen transfers from one domain to another. The boundaries between domains should be like airlocks, disinfecting people and objects that pass through them.

What protocols should these boundaries have? Since there are relatively few such crossings as compared to the number of patient contacts overall, we can consider more extensive decontamination procedures than just hand washing, such as clothing sanitation and the cleaning of cell phones and other personal effects. There is evidence that lab coats, PDAs, cell phones and the like act as repositories for pathogens, and can be responsible for HAI transmission. The protocol should still be efficient, and it can be. Staging such intensive interventions at the gateways could significantly reduce the flow of pathogens between patients.

\begin{figure}[h]
\centerline{
\includegraphics[width=175mm]{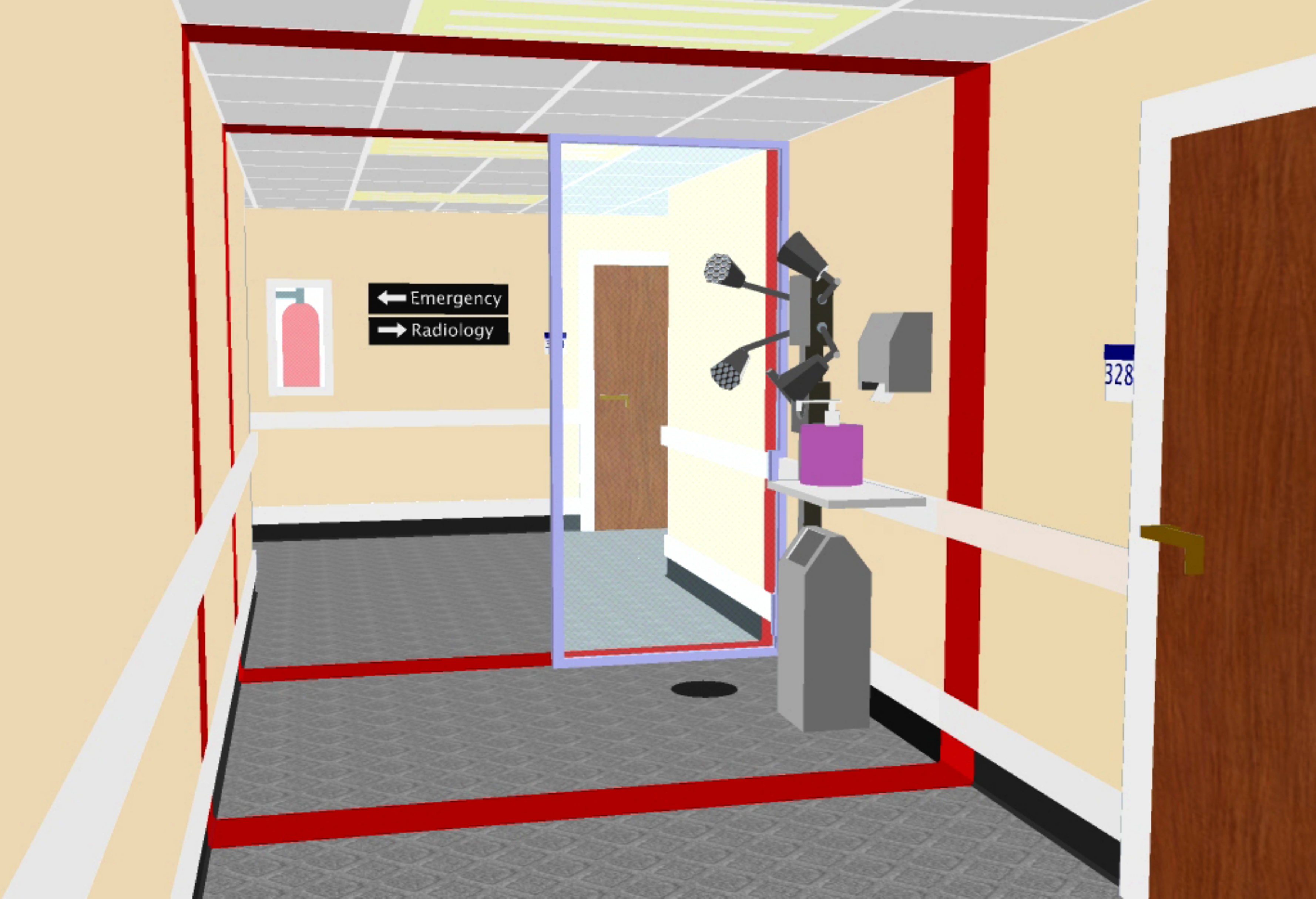}
}
\caption{\label{fig:airlock}A mockup of a disinfection gateway for use in inhibiting the spread of infections.}
\end{figure}

%
%
%
%

We have to consider how pathogens are transferred: from one patient to the surfaces and fabrics near that patient to the care providers and their clothing, cell phones and pagers. From there the pathogens are transferred either directly to another patient or to the fabrics and surfaces in common areas or around that patient from which they eventually reach that patient at a different contact opportunity.

Most of the possible transmission events happen because of the large number of contacts within a local ward between patients and doctors, nurses, medical technicians, food service people and cleaning staff. Each of these contacts has the potential to transfer pathogens between patients, and to contaminate objects in shared spaces, such as computer keyboards.

If there were no virulent pathogen in the ward in the first place, none of those possible transmission events could actually transfer virulent pathogens. Using boundary protocols to reduce transmission between wards would eliminate a large number of potential transmission events among the individuals within each ward.

With the use of boundary protocols, there would be a reduction not only in the transmission of existing pathogens but also in the emergence of new resistant strains. The high number of physical contacts makes medical care facilities a uniquely fertile environment for pathogens to evolve into more virulent strains. By blocking the spread of infection between areas, we can cut down on the appearance of virulent pathogens as well as their prevalence.

Would everyone have to go through disinfection at these airlocks? Visitors and patients entering a hospital for an appointment don't present the same level of risk (though they might be tested for infection themselves). Unlike care providers who go from patient to patient to patient, they don't act as agents for transmission. Accordingly, the same protocols need not apply. Similarly, a caregiver who is only interacting with a single patient need not undergo this process. Furthermore, these protocols could be overridden for the sake of speed in the event of an emergency---when protocols are generally observed, a single contact is unlikely to transmit pathogens.

The same principles of containment are behind biological membranes that prevent transmissions between parts of the body, and are the reason why the immune system is concentrated in the high-speed transport system of the body---the blood. It is the reason we have regulations about plant and animal products crossing national borders. Conversely, the absence of such boundary protections in an increasingly interconnected world has promoted the rise of highly virulent new strains of pathogens and the risks of global pandemics.

Reducing the probability of transmission at each provider-to-patient contact by hand washing and other protocols is still a good idea. At the same time, the flow of pathogens through a hospital and overall transmission between sites can be dramatically reduced. This can be done by creating additional levels of transmission-prevention at key internal boundaries in the care facility and between care facilities.

The cost of hospital-based infections is high and using high-leverage methods to eliminate them is the way to go. By instituting protocols at geographic domain boundaries, at low cost, we can dramatically reduce their transmission.

\section{Use E-Records for Research}
\label{sec:seven}

\begin{sectionsummary}
{\bf Scientific principle---``Big data'' research:} Our increasingly complex world yields massive quantities of data, and we now have the scientific knowledge to perform pattern recognition on the data. Scientists are utilizing such ``big data'' methods in areas as diverse as genomics, finance, and crime prevention. If made available, the vast corpus of medical records should result in the discovery of opportunities for advancement in medicine. This approach complements the more traditional and more controlled framework of specially designed clinical trials.
\end{sectionsummary}

Electronic records, which have become increasingly prevalent in recent years, represent a valuable repository of medical data. There are over 300 million people in the United States, most of whom are receiving some sort of medical care. If anonymized medical records were made available to researchers, these e-records could be leveraged to improve care at low cost.

In today's quest to answer questions about medicine and human health, the large-scale, controlled clinical trial is central. New drugs, surgical techniques, non-surgical interventions and medical devices are typically tested in such studies, which require the creation of control and test groups, controlling for confounding factors such as age and lifestyle, and the tracking of patients.

While these studies are essential for testing new drugs and interventions, not every medical question can or should be tested using a clinical trial, given the human and monetary resources that are required to conduct such studies.

Physicians and researchers already have other accepted ways of advancing medical knowledge.

Observational studies and chart reviews typically analyze groups of patients based on the condition displayed or the intervention used, and are often performed when large randomized studies are infeasible. They can yield important results even without the controls needed for clinical trials.

Physicians author case reports as a means of sharing their experiences with especially instructive cases. These reports are regarded as valuable parts of the medical literature and are a standard part of peer-reviewed medical journals. The medical field recognizes and accepts knowledge gained through observational studies, chart reviews and case reports in addition to controlled clinical studies.

With the use of e-records, the scope of chart reviews and the sharing of case reports are dramatically expanded by many orders of magnitude. This represents a unique opportunity to leverage vast amounts of newly available data to explore many medical questions.

Recently, the drug Vioxx was recalled due to side effects causing higher rates of heart attack than comparable drugs. The recall took place one month after results of a study using medical records from 1.4 million people were reported. This recall is one example of how collecting and using data from actual patients can lead to medical advances.

Another example of data gathering that has led to significant recent advances is the ongoing Framingham Heart Study. This longitudinal study has tracked over 10,000 individuals from three consecutive generations, monitoring their physical health and lifestyle choices, in order to learn about cardiovascular disease. That information has been used by researchers to make many advances, ranging from genetics to the role of social networks.

Many other studies have been based upon survey and medical reporting data collected by the CDC. The data that could be made available dwarfs current resources.

As previously noted, medical care data is logged for many millions of people in the United States. We can leverage the sheer volume of available data, combined with new pattern-recognition methods and theoretical advances in data analysis, to increase our knowledge in ways beyond the practical reach of other methods.

\begin{figure}[h]
\includegraphics{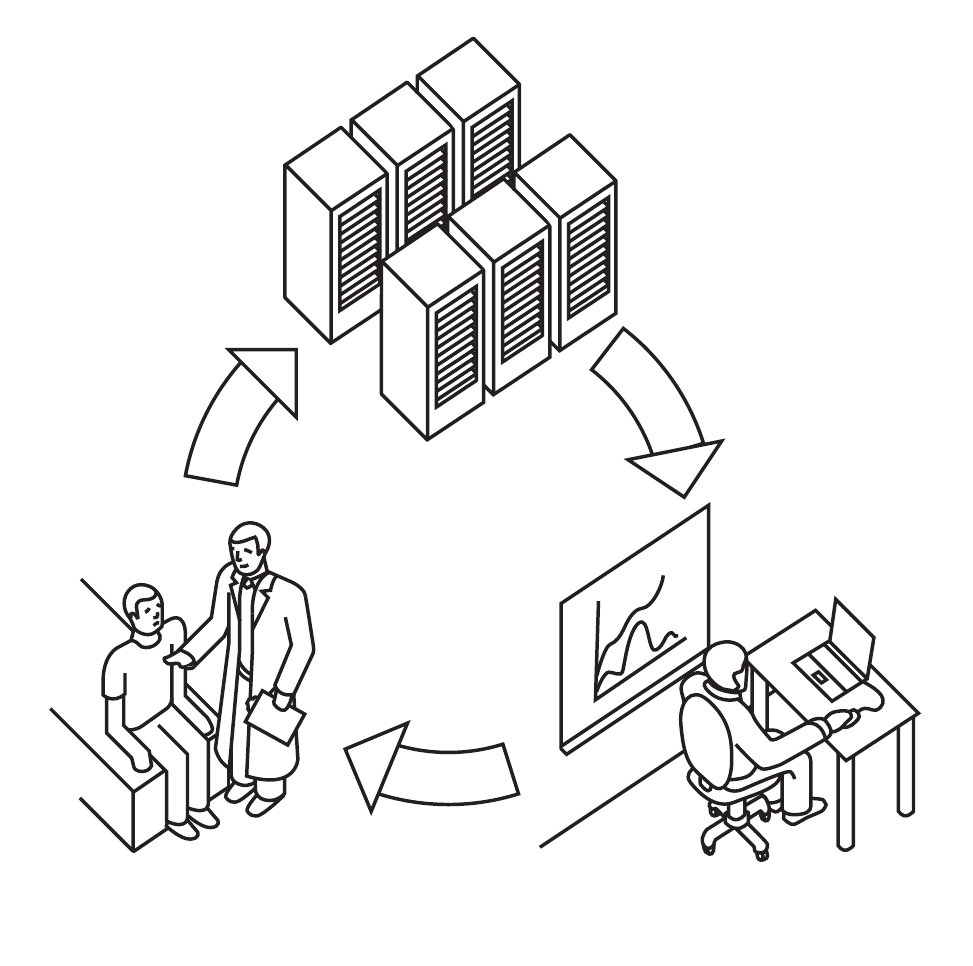}
\caption{\label{fig:records} Research based on medical care data can help improve healthcare.}
\end{figure}

At the very least, analyzing these data could reveal previously unknown connections---for example, between a particular medication, a bit of medical history, and a seemingly unrelated disease---that would provide clues as to what questions should be pursued in formal, controlled studies.

Also, since each person's medical records may cover many years, we can learn about long term effects much more easily and cost-effectively by analyzing these available data than by conducting longitudinal studies on a particular therapy. Thus, we can use these data to discover long-term effects that may otherwise not be detected at all.

Leveraging the availability of care data to increase our knowledge can't and shouldn't replace controlled studies or physician experience. But it can be a powerful and cost-effective tool, allowing us to utilize huge amounts of information and new methods of analysis to increase our medical knowledge, improving our ability to treat patients and take care of ourselves.

Data---be they from study results or individual experiences---are the raw material that we use to build our medical knowledge. Gaining access to such a huge volume of new information about human health is like inventing a microscope that can see objects that are much smaller, or a telescope that can see much farther away. This new data can lead to many new discoveries.

Analyses of these data would be an important addition to the medical research toolkit, augmenting traditional research methods. The governmental agencies that oversee various aspects of healthcare practice and research should work with medical organizations to make electronic medical records available in an anonymous, analyzable form. They should encourage use of this vast, important resource to propel our knowledge of medicine forward.

\section{Promote ``First Day'' Celebrations}
\label{sec:eight}

\begin{sectionsummary}
{\bf Scientific principle---Dynamics of collective behaviors:} Behavioral change can propagate through social network links.  The importance of individual behavioral choices to major public health problems and health more generally is well known. Individual responsibility for health can be socially reinforced. The power of social influence can be engaged to encourage individual healthy lifestyle choices. The most effective way to achieve a large response from a system is to engage its existing natural modes of activity. A yearly ``First Day'' celebration leverages the existing culture of ``New Year's resolutions'' and the natural yearly cycle of renewal.
\end{sectionsummary}

Any discussion about improving our healthcare system must acknowledge the important role that is played by people caring for themselves and their loved ones. The most important step we can take to improve the healthcare system is to support and inspire an informed and widespread level of personal care.

Major health issues are related to behavior---smoking, alcohol consumption, diet, exercise, even safe driving. Other health issues must be addressed partly through behavior, including remembering to take medications.

Addressing public health problems such as obesity is at times viewed as the responsibility of government, medical professionals or fast food chains. But these problems should also be addressed by individuals working to change their own behavior.

Yet, when we do turn to individuals to improve their own health habits, we often overlook the real potential in ensuring their success via support groups of friends and co-workers, and via support mechanisms such as community institutions and social traditions.

Al Gore's message calling for us all, as individuals and collectively, to be responsible for our planet, resonates in this instance. We can all take responsibility to safeguard our health. We need a culture of healthy people in a healthy world.

How do we realize this vision?

Building on the tradition of setting aside a time for New Year's resolutions, we can promote lifestyle change with the use of ``First Day'' celebrations, which will convey health information and will draw forth personal commitments to healthier living.

The fundamental purpose of these celebrations, resonating with ``today is the first day of the rest of your life'' is to celebrate healthy lifestyles for the new year. This will promote and reinforce our existing societal traditions and our recognition of the natural yearly cycle as one of renewal and improvement.

Health is serious business, but people should take care of their health in a positive way, mindful of new opportunities rather than focusing only on dangerous risks.

``First Day'' also builds on ``First Night,'' the popular New Year's Eve festivals full of arts, family activities and cultural entertainment. Started in 1976, First Night built upon people's natural tendency to celebrate the new year, and channeled that impulse toward constructive cultural activities and fun.

First Day should not be driven solely by individuals---companies, communities, towns, cities, and states can all play a role. The Centers for Disease Control and Prevention (CDC) articulates a vision of health as pervading all aspects of life. We can leverage personal and community participation to improve public health.

Perhaps surprisingly, Walmart has led the way. In 2007, Walmart launched a program in which employees design and carry out ``personal sustainability projects'' including anything from recycling at home to quitting smoking to getting more exercise. Originally focused on the environment, participants naturally included personal health projects. Indeed, health for oneself, one's family, community, country and world are all linked---both in effect and in desire and commitment for a better life.

Through this program, Walmart provides the framework for employees to exercise their capabilities. Working alongside others to accomplish goals has a positive effect on what people can accomplish. Employees self-monitor their progress for several weeks, and are encouraged to make the improvements long term. Co-workers encourage one another to meet their goals. Walmart's popular program has been a great success, helping many employees improve their lives.

This idea can be made into a national or global activity of personal and collective improvement. Aligning it with New Year's celebrations is a natural thing to do.

The preceding week, employers and government agencies can provide information and events. Organizations of different types---companies, religious organizations, schools, towns, states---can set up programs that encourage people to take responsibility for their own health and lifestyle, and they can provide supportive communities toward that end. The organizations themselves can undertake new commitments to improve social health and community well-being.

Some people may want their goals and commitments to be private or to share them with friends; others may be pleased to share them publicly. The key is for familiar institutions and networks to support each person's desire to improve his or her life and each person's journey toward better health.

Social network follow-up interactions can be planned. Internet-based and mobile device apps with calendars, reminders, and checklists can be developed to support people in reaching their goals.

We can dramatically improve health by inspiring individual responsibility and action. When people embrace their health as a personal opportunity and are also given community support, they reveal tremendous power to make lasting improvements in their own lives and each other's.

\section{References}

{\bf Introduction}
\begin{enumerate}
\item N. Kozhaya, Two myths about the American health care system. Montreal Economic Institute (2005).
\item World Health Statistics 2009. World Health Organization (2009).
\item B. Roehr, Health care in US ranks lowest among developed countries. {\it BMJ} {\bf 337}, a889 (2008).
\item S. C. Kulkarni, A. Levin-Rector, M. Ezzati, C. J. L. Murray. Falling behind: life expectancy in US counties from 2000 to 2007 in an international context. {\it Population Health Metrics} {\bf  9}, 16 (2011).
\item M. Z. Oestergaard, M. Inoue, S. Yoshida, W. R. Mahanani, F. M. Gore, S. Cousens, et al., Neonatal mortality levels for 193 countries in 2009 with trends since 1990: A systematic analysis of progress, projections, and priorities. {\it PLoS Medicine} {\bf  8}, 8 (2011).
\item Centers for Medicare and Medicaid Services, National Health Statistics Group, National health care expenditures data (2012).
\end{enumerate}

{\bf Section 1: Separate Simple Care}
\begin{enumerate}
\item Y. Bar-Yam, Improving the effectiveness of health care and public health: a multi-scale complex systems analysis. {\it American Journal of Public Health} {\bf 96},459-466 (2006).
\item Y. Bar-Yam, Making things work: solving complex problems in a complex world (NECSI Knowledge Press, Cambridge, MA, 2005). p. 239.
\item Y. Bar-Yam, Multiscale variety in complex systems. {\it Complexity} {\bf 9}, 37-45 (2004).
\item H. Baskas, More airports adding flu shots for fliers. {\it USA Today} (10/15/2008). Available from: \url{http://www.usatoday.com/travel/columnist/baskas/2008-10-14-flu-shots_N.htm}.
\item R. Bohmer, The rise of in-store clinics---threat or opportunity? {\it New England Journal of Medicine} {\bf 356}, 765-768 (2007).
\item J Groopman, How doctors think (Houghton Mifflin, New York, 2007).
\item P. H. Keckley, H. R. Underwood, M. Gandhi, Retail clinics: facts, trends, and implications; update and implications (Deloitte Center for Health Solutions, Washington DC, 2008).
\item J. Lambreaw, A Wellness Trust to prioritize disease prevention. Brookings Institution, report number: 2007-04 (2007).
\item M. Maciosek, A. Coffield, N. Edwards, T. Flottemesch, M. Goodman, L. Solberg, Priorities among effective clinical preventive services: results of a systematic review and analysis. {\it American Journal of Preventive Medicine} {\bf 31}, 52-61 (2006).
\item A. Mehrotra, H. Liu, J. L. Adams, M. C. Wang, J. R. Lave, N. M. Thygeson, L. I. Solberg, E. A. McGlynn, Comparing costs and quality of care at retail clinics with that of other medical settings for 3 common illnesses. {\it Annals of Internal Medicine} {\bf 151}, 321-328 (2009).
\item A. Mehrotra, M. C. Wang, J. R. Lave, J. L. Adams, E. A. McGlynn, Retail clinics, primary care physicians, and emergency departments: a comparison of patients' visits. {\it Health Affairs} {\bf 27}, 1272-1282 (2008).
\item Promoting healthy lifestyles (American Medical Association, Chicago, 2008).
\item R. Rudavsky, C. E. Pollack, A. Mehrotra, The geographic distribution, ownership, prices, and scope of practice at retail clinics. {\it Annals of Internal Medicine} {\bf 151}, 315-320 (2009).
\item M. K. Scott, Health care in the express lane: the emergence of retail clinics (California Healthcare Foundation, Oakland, CA, 2006).
\item M. Thygeson, K. A. Van Vorst, M. V. Maciosek, L. Solberg, Use and costs of care in retail clinics versus traditional care sites. {\it Health Affairs} {\bf 27}, 1283-1292 (2008).
\item U.S. Preventive Services Task Force Guide to Clinical Preventive Services. U.S. Department of Health and Human Services, Agency for Healthcare Research and Quality, report number: 07-05100 (2007).
\item K. Yarnall, K. Pollak, T. Ostbye, K. Krause, J. Michener, Primary care: Is there enough time for prevention? {\it American Journal of Public Health} {\bf 93}, 635-641 (2003). 
\end{enumerate}

{\bf Section 2: Empower Workgroup Competition}
\begin{enumerate}
\item K. J. Arrow, Uncertainty and the welfare economics of medical care. {\it American Economic Review}1963;53:941-973.
\item Y. Bar-Yam, Making things work (NECSI Knowledge Press, Cambridge, MA, 2005). See p. 239. 
\item M. Berwick, B. James, M. J. Coye, Connections between quality measurement and improvement. {\it Medical Care} {\bf 41}, 30-38 (2003).
\item B. Carrol, S. Tomas, Team competition spurs continuous improvement at Motorola. {National Productivity Review} {\bf 14}, 1-9 (1963).
\item D. Forsyth, Group dynamics (Wadsworth, Cengage Learning, Belmont, CA, 2006).
\item D. W. Johnson, G. Maruyama, R. Johnson, D. Nelson, L. Skon, Effects of cooperative, competitive, and individualist goal structures on achievement: A meta-analysis. {\it Psychological Bulletin} {\bf 89}, 47-62 (1981).
\item P. Krugman, Why markets can't cure healthcare. {\it New York Times} (7/25/2009). Available from: \url{http://krugman.blogs.nytimes.com/2009/07/25/why-markets-cant-cure-healthcare/}.
\item D. Levi, Group dynamics for teams (Sage Publications, Inc., Thousand Oaks, 2007).
\item E. A. McGlynn, Selecting common measures of quality and system performance. {\it Medical Care} {\bf 41}, 39-47 (2003).
\item D. A. Nadler, The effects of feedback on task group behavior: A review of the experimental research. {\it Organizational Behavior and Human Performance} {\bf 23}, 309-338 (1979).
\item P. Plsek,  Redesigning health care with insights from the science of complex adaptive systems. In: Institute of Medicine. Crossing the quality chasm: A new health system for the 21st century (National Academy Press, Washington DC, 2011).
\item M. E. Porter, E. O. Teisberg, Redefining health care: Creating value-based competition on results (Harvard Business School Press, Boston, 2006).
\item D. Schon, C. Argyris, Organizational learning II: Theory, method and practice (Addison-Wesley, Reading, MA, 1996).
\item D. Schon, C. Argyris, Organizational learning: A theory of action perspective (Addison-Wesley, Reading, MA, 1978).
\item P. M. Senge, The fifth discipline: The art and practice of the learning organization (Doubleday, New York, 1990).
\item F. E. Szarka, K. P. Grant, W. P. Flannery, Encouraging organizational learning through team competition. {\it Engineering Management Journal} {\bf 16}, 21-31 (2004).
\item M. Werner, D. A. Asch, The unintended consequences of publicly reporting quality information. {\it Journal of the American Medical Association} {\bf 293}, 1239-1244 (2008).
\end{enumerate}

{\bf Section 3: Create Superdoctor Teams}
\begin{enumerate}
\item D. Burke, H. Herrman, M. Evans, A. Cockram, T. Trauer, Educational aims and objectives for working in multidisciplinary teams. {\it Australasian Psychiatry} {\bf 8}, 336-339 (2000).
\item K. Calman, D. Hine, A policy framework for commissioning cancer services: A report by the expert advisory group on cancer to the chief medical officers of England and Wales (Department of Health, London, 1995).
\item J. H. Chang, E. Vines, H. Bertsch, D. L. Fraker, B. J. Czerniecki, E. F. Rosato, T. Lawton, E. F. Conant, S. G. Orel, L. Schuchter, K. R. Fox, N. Zieber, J. H. Glick, L. J. Solin, The impact of a multidisciplinary breast cancer center on recommendations for patient management. {\it Cancer} {\bf 91}, 1231-1237 (2001).
\item M. A. Denvir, J. P. Pell, A. J. Lee, J. Rysdale, R. J. Prescott, H. Eteiba, A. Walker, P. Mankad, I. R. Starkey, Variations in clinical decision-making between cardiologists and cardiac surgeons; a case for management by multidisciplinary teams? {\it J Cardiothoracic Surgery} {\bf 1} (2006).
\item A. Fendrick, R. Hirth, M. Chernew, Differences between generalist and specialist physicians regarding Helicobacter pylori and peptic ulcer disease. {\it American Journal Gastroenterology} {\bf 91}, 1544Ð1548 (1996).
\item A. Fleissig, V. Jenkins, S. Catt, L. Fallowfield, Multidisciplinary teams in cancer care: are they effective in the UK? {\it Lancet Oncology} {\bf 7}, 935Ð43 (2006).
\item D. Forsyth, Group dynamics.  Belmont CA: Wadsworth, Cengage Learning; 2006.
\item K. Gottlieb, I. Sylvester, D. Eby, Transforming your practice: what matters most. {\it Family Practice Management} {\bf 15,} 32-38 (2008).
\item L. R. Harrold, T. S. Field, J. H. Gurwitz, Knowledge, patterns of care, and outcomes of care for generalists and specialists. {\it Journal of General Internal Medicine} {\bf 14}, 499-511 (1999).
\item R. A. Hayward, The CalmanÐHine report: a personal retrospective on the UK's first comprehensive policy on cancer services. {\it Lancet Oncology} {\bf 7}, 336?346 (2006).
\item J. Hearn, I. J. Higginson, Do specialist palliative care teams improve outcomes for cancer patients? A systematic literature review. {\it Palliative Medicine} {\bf 12}, 317-332 (1998).
\item D. Levi, Group dynamics for teams (Sage Publications, Inc., Thousand Oaks, CA, 2007).
\item Medical Univerity of South Carolina, Hollings Cancer Center. Multidisciplinary team members [Internet]. 2010 [cited 2012 Jan 9]; Available from: http://www.muschealth.com/cancer/headandneckcancer/members.htm.
\item P. M. Senge, The fifth discipline: The art and practice of the learning organization (Doubleday, New York, 1990).
\item B. Starfield, L. Shi, J. Macinko, Contribution of primary care to health systems and health. {\it Milbank Quarterly} {\bf 83}, 457-502 (2005).
\item V. E. Stone, F. F. Mansourati, R. M. Poses, K. H. Mayer, Relation of physician specialty and HIV/AIDS experience to choice of guideline-recommended antiretroviral therapy. {\it Journal of General Internal Medicine} {\bf 16}, 360-368 (2001).
\end{enumerate}

{\bf Section 4: Accelerate Intake Routing}
\begin{enumerate}
\item C. Arnst, A new practice: the doctor will see you today. {\it Kaiser Health News} (7/14/2010). Available from: \url{http://www.kaiserhealthnews.org/Daily-Reports/2010/July/14/1khnstory.aspx?print=1}.
\item T. Bodenheimer, B. Lo, L. Casalino, Primary care physicians should be coordinators, not gatekeepers. {\it Journal of the American Medical Association} {\bf 281}, 2045-2049 (1999).
\item C. B. Forrest, Primary care gatekeeping and referrals: effective filter or failed experiment? {\it BMJ} {\bf 326}, 692-695 (2003).
\item S. Gatley, A. Grace, V. Lopes, E-referral and e-triage as mechanisms for enhancing and monitoring patient care across the primary-secondary provider interface. {\it Journal of Telemedicine and Telecare} {\bf 9}, 350-353 (2003). 
\item K. Grumbach, D. Keane, A. Bindman, Primary care and public emergency department overcrowding. {\it American Journal of Public Health} {\bf 83}, 372-378 (1993).
\item K. Grumbach, J. V. Selby, C. Damberg, A. B. Bindman, C. Quesenberry, A. Truman, C. Uratsu, Resolving the gatekeeper conundrum: what patients value in primary care and referrals to specialists. {\it Journal of the American Medical Association} {\bf 282}, 261-266 (1999).
\item V. Lattimer, A. George, F. Thompson, E. Thomas, M. Mullee, J. Turnbull, H. Smith, M. Moore, H. Bond, A. Glasper, Safety and effectiveness of nurse telephone consultation in out of hours primary care: randomised controlled trial. {\it BMJ} {\bf 317}, 1054-1059 (1998).
\item R. Leibowitz, S. Day, S. Dunt, A systematic review of the effect of different models of after-hours primary medical care services on clinical outcome, medical workload, and patient and GP satisfaction. {\it Family Practice} {\bf 20}, 311-317 (2003).
\item Merritt Hawkins \& Associates. 2009 survey of physician appointment wait times (MH\&A, Texas, 2009).
\item C. A. Moyer, D. T. Stern, D. S. Dobias, D. T. Cox, S. J. Katz, Bridging the electronic divide: patient and provider perspectives on e-mail communication in primary care. {\it American Journal of Managed Care} {\bf 8}, 427-433 (2002).
\item M. Murray, C. Tantau, Redefining open access to primary care. {\it Managed Care Quarterly} {\bf 7}, 45-55 (1999).
\item S. R. Poole, B. D. Schmitt, T. Carruth, A. Peterson-Smith, M. Slusarski, After- hours telephone coverage: the application of an area-wide telephone triage and advice system for pediatric practices. {\it Pediatrics} {\bf 92}, 670-679 (1993).
\item R. Richards, M. L. Navarro, R. W. Derlet, Survey of directors of emergency departments in California on overcrowding. {\it Western Journal of Medicine} {\bf 172}, 385Ð388 (2000).
\end{enumerate}

{\bf Section 5: Improve Communication}
\begin{enumerate}
\item L. Alvey, FDA and ISMP launch campaign to reduce medication mistakes caused by unclear medical abbreviations [press release]. Food and Drug Administration (6/14/2006).
\item E. Ammenwerth, P. Schnell-Inderst, C. Machan, U. Siebert, The effect of electronic prescribing on medication errors and adverse drug events: a systematic review. {\it Journal of the American Medical Informatics Association} {\bf 15}, 585-600 (2008).
\item J. S. Ash, D. F. Sittig, E. G. Poon, K. Guappone, E. Campbell, R. H. Dykstra, The extent and importance of unintended consequences related to computerized provider order entry. {\it Journal of the American Medical Informatics Association} {\bf 14}, 415 (2007).
\item Y. Bar-Yam, M. Smith, A. Wachman, S. Topolski, Prescription form with redundancy, version 0.5. New England Complex Systems Institute (8/26/2010). Available from: \url{http://necsi.edu/research/management/health/prescription.html}.
\item Y. Bar-Yam, Making things work (NECSI Knowledge Press, Cambridge, MA, 2004). See chapter 11.
\item Y. Bar-Yam, System care: multiscale analysis of medical errors---eliminating errors and improving organizational capabilities. New England Complex Systems Institute, report number 09-01-2004 (2004).
\item P. Bonnabry, C. Despont-Gros, D. Grauser, P. Casez, M. Despond, D. Pugin, C. Rivara-Mangeat, M. Koch, M. Vial, A. Iten, C. Lovis, A risk analysis method to evaluate the impact of a computerized provider order entry system on patient safety. {\it Journal of the American Medical Informatics Association} {\bf 15}, 453-460 (2008).
\item S. Borden, Computer entry a leading cause of medication errors in U.S. health systems: percentage of reported errors steadily increased from 1999 to 2003 [press release]. United States Pharmacopeia (12/20/2004).
\item B. Chaudhry, J. Wang, S. Wu, M. Maglione, W. Mojica, E. Roth, S. C. Morton, P. G. Shekelle, Systematic review: Impact of health information technology on quality, efficiency, and costs of medical care. {\it Annals of Internal Medicine} {\bf 144}, 742-752 (2006).
\item C. K. Christian, M. L. Gustafson, E. M. Roth, T. B. Sheridan, T. K. Gandhi, K. Dwyer, M. J. Zinner, M. M. Dierks, A prospective study of patient safety in the operating room. {\it Surgery} {\bf 139}, 159-173 (2006).
\item K. Colpaert, B. Claus, A. Somers, K. Vandewoude, H. Robays, J. Decruyenaere, Impact of computerized physician order entry on medication prescription errors in the intensive care unit: a controlled cross-sectional trial. {\it Critical Care} {\bf 10}, R21 (2006).
\item Experts to probe factors behind overdose error. {\it Edmonton Journal} (5/8/2007). Available from: \url{http://www.canada.com/edmontonjournal/news/story.html?id=adc17250-f77b-4b8f-b97f-d6657fe12e8d}.
\item Food and Drug Administration. Medication errors [Internet]. (Cited 1/9/2012). Available from: \url{http://www.fda.gov/drugs/drugsafety/medicationerrors/default.htm}.
\item A. Goldstein, Overdose kills girl at Children's Hospital. {\it Washington Post} (4/20/2001). Available from: \url{http://www.highbeam.com/doc/1P2-439883.html}.
\item M. Graban, Statistics on healthcare quality and patient safety [Weblog]. Leanblog (8/9/2009, cited 1/9/2012). Available from: \url{http://www.leanblog.org/2009/08/statistics-on-healthcare-quality-and/}.
\item R. W. Hamming, Error detecting and error correcting codes. {\it Bell Systems Technical Journal} {\bf 29}, 147-160 (1950).
\item Y. Y. Han, J. A. Carcillo, S. T. Venkataraman, R. S. B. Clark, R. S. Watson, T. C. Nguyen, H. Bayir, R. A. Orr, Unexpected increased mortality after implementation of a commercially sold computerized physician order entry system. {\it Pediatrics} {\bf 116}, 1506-1512 (2005).
\item T. Holdsworth, R. E. Fichtl, D. W. Raisch, A. Hewryk, M. Behta, E. Mendez-Rico, C. L. Wong, J. Cohen, S. Bostwick, B. M. Greenwald, Impact of computerized prescriber order entry on the incidence of adverse drug events in pediatric inpatients. {\it Pediatrics} {\bf 120}, 1058-1066 (2007).
\item Institute of Medicine. Preventing medication errors: quality chasm series (National Academies Press, Washington DC, 2007).
\item Institute of Medicine. To err is human: building a safer health system (The National Academies Press, Washington DC, 2000).
\item A. G. Kennedy, B. Littenberg, A modified outpatient prescription form to reduce prescription errors. {\it Joint Commission Journal on Quality and Patient Safety} {\bf 30}, 480-487 (2004).
\item R. Koppel, J. P. Metlay, A. Cohen, B. Abaluck, A. R. Localio, S. E. Kimmel, B. L. Strom, Role of computerized physician order entry systems in facilitating medication errors. {\it Journal of the American Medical Association} {\bf 293}, 1197-1203 (2005).
\item L .L. Leape et. al., Systems analysis of adverse drug events. {\it Journal of the American Medical Association} {\bf 274}, 35-43 (1995).
\item L. L. Leape, Errors in medicine. {\it Journal of the American Medical Association} {\bf 272}, 1851-1857 (1994).
\item D. Liebovitz, Health care information technology: A cloud around the silver lining? {\it Archives of Internal Medicine} {\bf 169}, 924-926 (2009).
\item S. Loughran, In-hospital deaths from medical errors at 195,000 per year, HealthGrades study finds [press release]. HealthGrades (7/27/2004).
\item F. J. MacWilliams, N. J. A. Sloane, The theory of error-correcting codes (Elsevier B.V., Amsterdam, 1977).
\item C. Perrow, Normal accidents: living with high risk technologies (Princeton University Press, Princeton, NJ, 1999).
\item Quality Interagency Coordination Task Force. Doing what counts for patient safety: federal actions to reduce medical errors and their impact. Report to the president on medical errors (2/2000).
\item Quality Interagency Coordination Task Force. National summit on medical errors and patient safety research, summary [Internet] (11/20/2006, cited 1/9/2012). Available from: \url{http://www.quic.gov/summit/summary1.htm}.
\item J. Reason, Human error: models and management. {\it BMJ} {\bf 320}, 768Ð770 (2000).
\item A. Robeznieks, Data entry is a top cause of medication errors: training and design are seen as keys to reducing electronic prescribing errors. {\it American Medical News} (1/24/2005). Available from: \url{http://www.ama-assn.org/amednews/2005/01/24/prsa0124.htm}.
\item C. E. Shannon, A mathematical theory of communication, {\it Bell Systems Technical Journal} {\bf 27}, 379Ð423, 623Ð656 (1948).
\item R. Shapiro, Preventable medical malpractice: revisiting the Dennis Quaid medication/hospital error case [Weblog]. {\it The Injury Board Blog Network} (8/9/2010, cited 1/9/2012). Available from: \url{http://virginiabeach.injuryboard.com/medical-malpractice/preventable-medical-malpractice-revisiting-the-dennis-quaid-medicationhospital-error-case.aspx}.
\item C. L. Streeter, Redundancy in organizational systems. {\it The Social Services Review} {\bf 66}, 97-111 (1992).
\item J. M. Teich, P. R. Merchia, J. L. Schmiz, G. J. Kuperman, C. D. Spurr, D. W. Bates, Effects of computerized physician order entry on prescribing practices. {\it Archives of Internal Medicine} {\bf 160}, 2741-2747 (2000).
\item US Dept. of Health and Human Services, Agency for Healthcare Research and Quality. Computerized provider order entry [Internet]. (Cited 1/9/2012). Available from: \url{http://psnet.ahrq.gov/primer.aspx?primerID=6}.
\item K. E. Walsh, W. G. Adams, H. Bauchner, R. J. Vinci, J. B. Chessare, M. R. Cooper, P. M. Hebert, E. G. Schainker, C. P. Landrigan, Medication errors related to computerized order entry for children. {\it Pediatrics} {\bf 118}, 1872-1879 (2006).
\item R. L. Wears, M. Berg, Computer technology and clinical work: still waiting for Godot. {\it Journal of the American Medical Association} {\bf 293}, 1261-1263 (2005).
\item S. J. Weiner, A. Schwartz, F. Weaver, J. Goldberg, R. Yudkowsky, G. Sharma, A. Binns-Calvey, B. Preyss, M. M. Schapira, S. D. Persell, E. Jacobs, R. I. Abrams, Contextual errors and failures in individualizing patient care: a multicenter study. {\it Annals of Internal Medicine} {\bf 153}, 69-75 (2010).
\end{enumerate}

{\bf Section 6: Create Disinfection Gateways}
\begin{enumerate}
\item Agency for Healthcare Research and Quality. Health care-associated infections [Internet]. (Cited 1/9/2012). Available from: \url{http://psnet.ahrq.gov/primer.aspx?primerID=7}.
\item B. McCaughey, Unnecessary deaths: the human and financial costs of hospital infections, 3rd edition (Committee to Reduce Infection Deaths, New York, 2006).
\item C. Backman, D. E. Zoutman, P. B. Marck, An integrative review of the current evidence on the relationship between hand hygiene interventions and the incidence of health care-associated infections. {\it American Journal of Infection Control} {\bf 36}, 333-348 (2008).
\item C. A. Bascetta, N. Edwards, D. Brown, E. Peterson, A. E. Richardson, S. S. Legeer, et al., Health-care-associated infections in hospitals: an overview of state reporting programs and individual hospital initiatives to reduce certain infections. U. S. Government Accountability Office, report number: GAO-08-808 (2008).
\item J. M. Boyce, D. Pittet, Guideline for hand hygiene in health-care settings: recommendations of the healthcare infection control practices advisory committee and the HICPAC/SHEA/APIC/IDSA Hand Hygiene Task Force. {\it Infection Control and Hospital Epidemiology} {\bf 23}, S3-40 (2002).
\item J. M. Boyce, Environmental contamination makes an important contribution to hospital infection. {\it Journal of Hospital Infection} {\bf 65,} S2,50-54 (2007).
\item C. M. Braddy, J. E. Blair, Colonization of personal digital assistants used in a health care setting. {\it American Journal of Infection Control} {\bf 33}, 230-232 (2005).
\item R. R. W. Brady, S. F. Fraser, M. G. Dunlop, S. Paterson-Brown, A. P. Gibb, Bacterial contamination of mobile communication devices in the operative environment. {\it Journal of Hospital Infection} {\bf 66}, 397-398 (2007). 
\item R. R. W. Brady, J. Verran, N. N. Damani, A. P. Gibb, Review of mobile communication devices as potential reservoirs of nosocomial pathogens. {\it Journal of Hospital Infection} {\bf 71}, 295-300 (2009).
\item S. Bures, J. T. Fishbain, C. F. T. Uyehara, J. M. Parker, B. W. Berg, Computer keyboards and faucet handles as reservoirs of nosocomial pathogens in the intensive care unit. {\it American Journal of Infection Control} {\bf 28}, 465-471 (2000).
\item Center for Disease Control. Monitoring hospital-acquired infections to promote patient safety---United States, 1990Ð1999. {\it Morbidity and Mortality Weekly Report} {\bf 49}, 149-153 (2000).
\item Centers for Disease Control and Prevention. Healthcare-associated infections (HAIs) [Internet]. (Cited 1/9/2012). Available from: \url{http://www.cdc.gov/hai/}.
\item T. Donker, J. Wallinga, H. Grundmann, Patient referral patterns and the spread of hospital-acquired infections through national health care networks. {\it PLoS Computational Biology} {\bf 6}, e1000715 (2010).
\item N. Duckro, D. W. Blom, E. A. Lyle, R. A. Weinstein, M. K. Hayden, Transfer of vancomycin-resistance enterococci via health care worker hands. {\it Archives of Internal Medicine} {\bf 165}, 302-307 (2005).
\item M. R. Eber, R. Laxminarayan, E. N. Perencevich, A. Malani, Clinical and economic outcomes attributable to health care-associated sepsis and pneumonia. {\it Archives of Internal Medicine} {\bf 170}, 347-353 (2010).
\item T. Eckmanns, F. Schwab, J. Bessert, R. Wettstein, M. Behnke, H. Grundmann, H. Ruden, P. Gastmeir, Hand rub consumption and hand hygiene compliance are not indicators of pathogen transmission in intensive care units. {\it Journal of Hospital Infection} {\bf 63}, 406-411 (2006).
\item B. C. Eckstein, D. A. Adams, E. C. Eckstein, A. Rao, A. K. Sethi, G. K. Yadavalli, C. J. Donskey, Reduction of Clostridium difficile and vancomycin- resistant Enterococcus contamination of environmental surfaces after an intervention to improve cleaning methods. {\it BMC Infectious Diseases} {\bf 7}, 61 (2007).
\item T. G. Emori, D. H. Culver, T. C. Horan, W. R. Jarvis, J. W. White, D. R. Olson, S. Banerjee, J. R. Edwards, W. J. Martone, R. P. Gaynes, National nosocomial infections surveillance system (NNIS): description of surveillance methods. {\it American Journal of Infection Control} {\bf 19}, 19-35 (1991).
\item S. Fridkin, S. Welbel, R. Weinstein, Magnitude and prevention of nosocomial infections in the intensive care unit. {\it Infectious Disease Clinics of North America} {\bf 11}, 479-496 (1997).
\item Institute for Healthcare Improvement. Reducing healthcare-associated infections [Internet]. (2011, cited 1/9/2012). Available from: \url{http://www.ihi.org/explore/HAI/Pages/default.aspx}. 
\item R. M. Klevens, J. R. Edwards, C. L. Richards Jr., T. C. Horan, R. P. Gaynes, D. A. Pollock, et al., Estimating health care-associated infections and deaths in U.S. hospitals, 2002. National Center for Infectious Diseases, Center for Disease Control and Prevention, report number: 122 (2007). See pg. 160-166.
\item A. Kramer, I. Schwebke, G. Kampf, How long do nosocomial pathogens persist on inanimate surfaces? A systematic review. {\it BMC Infectious Diseases} {\bf 6}, 130 (2006).
\item L. Leykum, Y. Bar Yam, The rational for system level strategies of infection control. New England Complex Systems Institute. Report number 08-17-2010 (2010).
\item M. Lipsitch, C. T. Bergstrom, B. R. Levin, The epidemiology of antibiotic resistance in hospitals: paradoxes and prescriptions. {\it PNAS} {\bf 97}, 1938-1943 (2000).
\item K. K. Macartney, M. H. Gorelick, M. L. Manning, R. L. Hodinka, L. M. Bell, Nosocomial respiratory syncytial virus infections: the cost-effectiveness and cost-benefit of infection control. {\it Pediatrics} {\bf 106}, 520-526 (2000).
\item B. McCaughey, Hospital scrubs are a germy, deadly mess: bacteria on doctor uniforms can kill you. {\it Wall Street Journal} (1/8/2009). Available from: \url{http://online.wsj.com/article/SB123137245971962641.html}.
\item L. C. McDonald, W. R. Jarvis, Linking antimicrobial use to nosocomial infections: the role of a combined laboratory-epidemiology approach. {\it Annals of Internal Medicine} {\bf 129}, 245-247 (1998).
\item C. A. Muto, J. A. Jernigan, B. E. Ostrowsky, H. M. Richet, W. R. Jarvis, J. M. Boyce, B. M. Farr, SHEA Guideline for Preventing Nosocomial Transmission of Multidrug-Resistant Strains of Staphylococcus aureus and Enterococcus. {\it Infection Control and Hospital Epidemiology} {\bf 24}, 362-386 (2003).
\item N. P. O'Grady, M. Alexander, E. P. Dellinger, J. L. Gerberding, S. O. Heard, D. G. Maki, H. Masur, R. D. McCormick, L. A. Mermel, M. L. Pearson, I. I. Raad, A. Randolph, R. A. Weinstein, Guidelines for the prevention of intravascular catheterÐrelated infections. {\it Infection Control and Hospital Epidemiology} {\bf 23}, 759-769 (2002).
\item D. Pittet, S. Hugonnet, S. Harbarth, P. Mourouga, V. Sauvan, S. Touveneau, T. V. Perneger, Effectiveness of a hospital-wide programme to improve compliance with hand hygiene. {\it Lancet} {\bf 356}, 1307Ð12 (2000).
\item E. M. Rauch, Y. Bar-Yam, Long-range interactions and evolutionary stability in a predator-prey system. {\it Physical Review E} {\bf 73}, 020903 (2006).
\item R. D. Scott II, The direct medical costs of healthcare-associated infections in U.S. hospitals and the benefits of prevention. Centers for Disease Control and Prevention. 2009 Jan.
\item V. SŽbille, S. Chevret, A. Valleron, Modeling the spread of resistant nosocomial pathogens in an intensive care unit. {\it Infection Control and Hospital Epidemiology} {\bf 18}, 84-92 (1997).
\item V. SŽbille, A. Valleron, A computer simulation model for the spread of nosocomial infections caused by multidrug-resistant pathogens. {\it Computational and Biomedical Research} {\bf 30}, 307-322 (1997).
\item L. Silvestri, A. J. Petros, R. E. Sarginson, M. A. de la Cal, A. E. Murray, H. K. F. van Saene, Handwashing in the intensive care unit: a big measure with modest effects. {\it Journal of Hospital Infection} {\bf 59}, 172-9 (2005).
\item C. Steere, G. F. Mallison, Handwashing practices for the prevention of nosocomial infections. {\it Annals of Internal Medicine} {\bf 83}, 683-690 (1975).
\item The Department of Health and Human Services. HHS Action plan to prevent healthcare-associated infections [Internet]. (Cited 1/9/2012). Available from: \url{http://www.hhs.gov/ash/initiatives/hai/actionplan/}.
\item A. M. Treakle, K. A. Thorn, J. P. Furuno, S. M. Strauss, A. D. Harris, E. N. Perencevich, Bacterial contamination of health care workers' white coats. {\it American Journal of Infection Control} {\bf 37}, 101-105 (2009).
\item R. Wenzel, The economics of nosocomial infections. {\it Journal of Hospital Infection} {\bf 31}, 79-87 (1995).
\item W. W. Williams, CDC guideline for infection control in hospital personnel. {\it Infection Control} {\bf 4}, S326-349 (1983).
\item World Health Organization. Prevention of hospital-acquired infections: a practical guide, 2nd edition. (WHO, Malta, 2002).
\item D. S. Yokoe, L. A. Mermel, D. J. Anderson, et al., A compendium of strategies to prevent healthcare-associated infections in acute care hospitals. Supplement article: executive summary. {\it Infection Control and Hospital Epidemiology} {\bf 29}, S12 (2008).
\end{enumerate}

{\bf Section 7: Use E-Records For Research}
\begin{enumerate}
\item J. J. Allison, T. C. Wall, C. M. Spettell, J. Calhoun, C. A. Fargason, R. Kobylinski, R. Farmer, C. I. Kiefe, The art and science of chart review. {\it Joint Comm J Qual Improv} {\bf 26}, 115-36 (2000).
\item D. G. Altman, K. F. Schulz, D. Moher, M. Egger, F. Davidoff, D. Elbourne, P. C. G¿tzsche, T. Lang, The revised CONSORT statement for reporting randomized trials: explanation and elaboration. {\it Annals of Internal Medicine.} {\bf 134}, 663-694 (2001).
\item K. Benson, A. J. Hartz, A comparison of observational studies and randomized, controlled trials. {\it New England Journal of Medicine} {\bf 342}, 1878-1886 (2000).
\item N. Black, Why we need observational studies to evaluate the effectiveness of health care. {\it BMJ} {\bf 312}, 1215 (1996).
\item Centers for Disease Control and Prevention. Data \& Statistics [Internet]. (Cited 1/9/2012). Available from: \url{http://www.cdc.gov/datastatistics/}.
\item Centers for Disease Control and Prevention. National Center for Health Statistics [Internet]. (Cited 1/9/2012). Available from: \url{http://www.cdc.gov/nchs/}.
\item Centers for Disease Control and Prevention. WONDER online databases [Internet]. (Cited 1/9/2012). Available from: \url{http://wonder.cdc.gov/}.
\item N. A. Christakis, J. H. Fowler, The spread of obesity in a large social network over 32 years. {\it New England Journal of Medicine} {\bf 357}, 370-379 (2007).
\item D. C. Des Jarlais, C. Lyles, N. Crepaz, TREND Group. Improving the reporting quality of nonrandomized evaluations of behavioral and public health interventions: the TREND statement. {\it American Journal of Public Health} {\bf 94}, 361-366 (2004).
\item Framingham Heart Study: A Project of National Heart, Lung, and Blood Institute and Boston University (1948-present) [Internet]. (1/3/2012, cited 1/9/2012). Available from: \url{http://www.framinghamheartstudy.org/}.
\item D. J. Graham, D. Campen, R. Hui, M. Spence, C. Cheetham, G. Levy, S. Shoor, W. A. Ray, Risk of acute myocardial infarction and sudden cardiac death in patients treated with cyclo-oxygenase 2 selective and non-selective non- steroidal anti-inflammatory drugs: nested case-control study. {\it Lancet} {\bf 365}, 475Ð81 (2005).
\item R. W. Haley, D. R. Schaberg, D. K. Mcclish, D. Quade, K. B. Crossley, D. H. Culver, W. M. Morgan, J. E. Mcgowan Jr., R. H. Shachtman, The accuracy of retrospective chart review in measuring nosocomial infection rates: results of validation studies in pilot hospitals. {\it American Journal of Epidemiology} {\bf 111}, 516-533 (1980).
\item M. Jenicek, Clinical case reporting in evidence-based medicine, 2nd edition (Hodder Arnold Publication, London, 2001).
\item M. Kidd, C. Hubbard, Introducing the Journal of Medical Case Reports. {\it Journal of Medical Case Reports} {\bf 1}, 1 (2007).
\item D. Levy, A. L. DeStefano, M. G. Larson, C. J. O'Donnell, R. P. Lifton, H. Gavras, L. A. Cupples, R. H. Myers, Evidence for a gene influencing blood pressure on chromosome 17: genome scan linkage results for longitudinal blood pressure phenotypes in subjects from the Framingham Heart Study. {\it Hypertension} {\bf 36}, 477 (2000).
\item G. Narcisi, Clinical data repositories: boosting patient care and research for a data-intensive future. {\it CMIO} 2010 Apr:8-9.
\item C. J. O'Donnell, K. Lindpaintner, M. G. Larson, V. S. Rao, J. M. Ordovas, E. J. Schaefer, R. H. Myers, D. Levy, Evidence for association and genetic linkage of the angiotensin-converting enzyme locus with hypertension and blood pressure in men but not women in the Framingham Heart Study. {\it Circulation} {\bf 97}, 1766-1772 (1998).
\item M. L. Richardson, F. S. Chew, Radiology Case Reports: a new peer-reviewed, open-access journal specializing in case reports. {\it Journal of Radiology Case Reports} {\bf 1}, 1-3 (2006).
\item R. Rubin, How did Vioxx debacle happen? {\it USA Today} (10/12/2004). Available from: \url{http://www.usatoday.com/news/health/2004-10-12-vioxx-cover_x.htm}.
\item R. Schoenberg, C. Safran, Internet based repository of medical records that retains patient confidentiality. {\it BMJ} {\bf 321}, 1199-1203 (2000).
\item A. Schwartz, C. Pappas, L. J. Sandlow, Data repositories for medical education research: issues and recommendations. {\it Academic Medicine} {\bf 85}, 837-843 (2010).
\item University of Virginia School of Medicine. Clinical Data Repository (CDR) [Internet]. (6/2/2010, cited 1/9/2012). Available from: \url{http://www.medicine.virginia.edu/clinical/departments/phs/informatics/cdr-page}.
\item C. G. Victora, J. Habicht, J. Bryce, Evidence-based public health: moving beyond randomized trials. {\it American Journal of Public Health} {\bf 94}, 400-405 (2004).
\item Wake Forest University School of Medicine, Clinical Data Repository (CDR).
\item E. R. Weitzman, L. Kacil, K. D. Mandl, Sharing medical data for health research: the early personal health record experience. {\it Journal of Medical Internet Research} {\bf 12}, e14 (2010).
\end{enumerate}

{\bf Section 8: Promote ÒFirst DayÓ Celebrations}
\begin{enumerate}
\item M. Barbaro, At Wal-Mart, lessons in self-help. {\it New York Times} (4/5/2007). Available from: \url{http://www.nytimes.com/2007/04/05/business/05improve.html?pagewanted=all}.
\item Centers for Disease Control and Prevention. CDC's healthy communities program. (9/23/2011, cited 1/9/2012). Available from: \url{http://www.cdc.gov/healthycommunitiesprogram/}.
\item M. Minkler, Personal responsibility for health? A review of the arguments and the evidence at century's end. {\it Health Education and Behavior} {\bf 26}, 121-141 (1999).
\item J. C. Norcross, M. S. Mrykalo, M. D. Blagys. Auld lang syne: success predictors, change processes, and self-reported outcomes of New Year's resolvers and nonresolvers. {\it Journal of Clinical Psychology} {\bf 58}, 397-405 (2002).
\item A. P. Perez, M. M. Phillips, C. E. Cornell, G. Mays, B. Adams, Promoting dietary change among state health employees in Arkansas through a worksite wellness program: the healthy employee lifestyle program (HELP). {\it Preventing Chronic Disease: Public Health Research, Practice, and Policy} {\bf 6}, A123 (2009).
\item The National Association of Community-Based New Year's Eve Arts Festivals. Welcome to First Night USA [Internet]. (Cited 1/9/2012). Available from: \url{http://firstnightusa.org/}.
\item G. Turner, Peer support and young people's health. {\it Journal of Adolescence} {\bf 22}, 567Ð572 (1999).
\item Wal-Mart announces expansion of associate-driven personal sustainability projects [press release] Wal-Mart (4/5/2007).
\item Wal-Mart. 2009 global sustainability report: associates' personal sustainability projects [Internet]. (Cited 1/9/2012). Available from: \url{http://walmartstores.com/sites/sustainabilityreport/2009/s_ao_psp.html}.
\item Wal-Mart. The personal sustainability project [Internet]. (Cited 1/9/2012). Available from: \url{http://walmartstores.com/sites/sustainabilityreport/2007/associatesPersonal.html}.
\item R. S. Zimmerman, C. Connor, Health promotion in context: the effects of significant others on health behavior change. {\it Health Education and Behavior} {\bf 16}, 57-75 (1989).
\end{enumerate}

\end{document}